\documentclass[onecolumn,aps,prd,notitlepage]{revtex4-1}
\usepackage{cancel}
\usepackage{graphicx}
\usepackage{color}
\usepackage{setspace}
\usepackage{fancyhdr}
\usepackage{amssymb}
\usepackage{amsmath}
\usepackage{float}
\usepackage{url}
\usepackage{todonotes}
\usepackage[utf8]{inputenc}

\usepackage[framemethod=default]{mdframed}
\newmdenv[linecolor=red,backgroundcolor=yellow]{myframe}

\newcommand{\bwt}{\begin{widetext}}
\newcommand{\ewt}{\end{widetext}}
\newcommand{\beq}{\begin{equation}}
\newcommand{\eeq}{\end{equation}}
\newcommand{\bea}{\begin{eqnarray}}
\newcommand{\eea}{\end{eqnarray}}

\begin{document}
\title{Gravitational quadrupole deformation and the tidal deformability for stellar systems: (The number of) Love for undergraduates} 
\author{Andreas Zacchi}
\affiliation{Institut f\"ur Theoretische Physik, Goethe Universit\"at, Max-von-Laue-Stra\ss{}e 1, D-60438 Frankfurt, Germany} 
 \date{\today}

\begin{abstract}
This article is intended for undergraduate students with the aim to provide a pedagogical introduction to
the physics of stellar tidal deformations. The spherically symmetric shape of any star is deformed
via rotation around an arbitrary axis or by the presence of an external tidal field. We compute the
ellipticity of such a stellar object and show that rotation can be treated analogously to tidal effects
caused by an external field, which induces a quadrupole moment. 
The detection of gravitational waves from a binary neutron star merger (GW170817) in 2017 set constraints on the tidal deformability parameter $\Lambda $ and on equations of state for compact stars. We derive the corresponding formalism classically and show that the Newtonian limit taken via general relativity is justified. The compressibility and the compactness $C=M/R$ of matter are discussed classically and compared to the relativistic approach. We find that relativistically the main influence on tidal deformability is related to the star’s compactness $C$. 
\end{abstract}
\maketitle
\section{Motivation}
\section{Introduction and Motivation}
The purpose and motivation of this article is to help undergraduate students and/or their supervisors to find a rather pedagogical approach to the study of tidal effects in stellar systems. Since the detection of gravitational waves from a binary black hole-black hole merger in 2015 \cite{Abbott:2016blz,Abbott:2016nmj,TheLIGOScientific:2016pea} and two years later from the neutron star merger event GW170817 \cite{TheLIGOScientific:2017qsa,Abbott:2018wiz,Annala:2017llu}, the
tidal deformability of compact stars is an area of physics of increasing popularity.\\ \\
The main focus of attention is laid upon a classical approach to the corresponding equations. It
is thereby intended to generate a progressive introduction to this area of physics and to gain a better understanding of the corresponding equations and related quantities. First it is shown that the rotation of a star around an arbitrary axis obeys in principle the same physics as the effect of an external tidal field. Our analysis of the quadrupole moment, the hydrostatic equilibrium and the corresponding Poisson equation yields a differential equation the solution of which resembles a Bessel function very closely. The evaluation of this equation eventually gives quantities such as the tidal Love number $k_2$, which
we can relate to the quadrupole parameter $J_2$ resulting from the classical evaluation of the quadrupole tensor. The classical derivation of this formalism yields the correct limit taken from general relativity which is used in several articles on the subject of a compact star’s tidal deformability \cite{Hinderer:2007mb,Flanagan:2007ix,Hinderer:2009ca,Postnikov:2010yn}.
Proceeding and expanding the classical formalism relativistically, we also show that for stars resulting from polytropic equations of state (EoS) and selfbound stars obtained with a constant speed of sound EoS, the mass-radius relations scale with a constant to some power. The resulting compactnesses $C=M/R$ and the tidal Love numbers $k_2$ are independent of these constants, so that constant factors in any EoS do not change the deformability properties. Classical results are compared with results from general relativity by solving the Tolman-Oppenheimer-Volkoff equations. The finding
that a compact star with a smaller compactness $C$ is more deformable compared to a more compact star is not surprising, but also the stiffness or softness of an EoS plays a nontrivial role. Several polytropic indices are hence studied and evaluated on their influence on the compressibility of the matter the star is made of. Classically $C = 0$ and the tidal deformability is described only by the Love number $k_2$. In general relativity  $C \neq 0$ and the compactness 
gives the main contribution, although $k_2$ is still of influence.

\section{Newtonian physics}
In 1867 Isaac Newton published his work entitled \textit{Philosophiae naturalis principia mathematica}. In this work he linked the laws of physics on earth to the laws of physics in space. Within this unified theory it is not only possible to derive the Keplerian laws but also to study stellar physics.
\subsection{Non-relativistic stellar structure equations}
Two forces act within an ordinary star. One of these forces is gravity $F_G$ and the other force arises from the pressure $p$ which counterbalances gravity to keep the star in an equilibrium state \cite{Silbar04,Sagert:2005fw}. The classical treatment due to Isaac Newton in infinitesimal calculus reads 
\begin{equation}\label{garmir_thehounddog}
 dF_G=\frac{-G dm M(r)}{r^2}
\end{equation}
where $G=6.67 \times 10^{-11} {\rm{m^3 kg^{-1} s^{-2}}}$ is the gravitational constant in SI units, $r$ the radial distance of a spherically symmetric star. Mass conservation is given by 
\begin{equation}\label{mass_appeal_madness}
 dm=4\pi r^2\rho(r)dr 
\end{equation}
with $\rho(r)$ being the mass density.  The pressure acting on a surface A is 
\begin{equation}\label{pressure_appeals_madness}
 dp=\frac{dF_G}{A}=\frac{dF_G}{4 \pi r^2}.
\end{equation}
Combining eqs.~(\ref{garmir_thehounddog}), (\ref{mass_appeal_madness}) and (\ref{pressure_appeals_madness}) eventually leads to 
\begin{equation}\label{hydro_schraubenschluessel}
\frac{dp}{dr}=-\frac{G M(r)\rho(r)}{r^2}.
\end{equation}
Eq.~(\ref{hydro_schraubenschluessel}) tells us that a sphere in hydrostatic equilibrium balances the graviational pull with the pressure of the matter pushing outwards. In case of ordinary stars such as our sun this counter pressure on gravity is of thermal origin and provided by fusion processes. In case of compact stars the counter pressure is arranged by the degeneracy pressure of the corresponding particles. In case of white dwarfs relativistic electrons arrange for an adequate counter pressure counterbalancing the pull of gravity. 
More compact stars such as neutron stars are stabilized by the counter pressure of relativistic neutrons and interactions among them. Since neutron stars are highly relativistic objects, eq.~(\ref{hydro_schraubenschluessel}) has to be adjusted. This is discussed in section \ref{gr_baby}.
\subsection{Polytropic Equation of state}\label{sadogoat}
In the last section we motivated differential equations for the determination of a star's mass (eq.~(\ref{mass_appeal_madness})) and it's corresponding pressure (eq.~(\ref{hydro_schraubenschluessel})) depending on the star's radius $r$. What we have to do now is to set up a relation between the pressure $p$ and the mass density $\rho(r)$. Such a relation is called the equation of state (EoS), or $p(\rho(r))$. Note that in the relativistic case the mass density $\rho$ is replaced by the energy density $\epsilon$ via $\rho(r)c^2=\epsilon(r)$, where c is the speed of light. 
\\ 
In this section we shortly sketch how to derive a polytropic EoS for compact stars according to \cite{Silbar04,Sagert:2005fw,Schmitt:2010pn} because we discuss such EoSs in the following.\\
In a quantum system $E_{kin.} \gg E_{therm.}$, so that degenerate fermions can be described at $T=0$ due to the Pauli principle.  The energy to enter the system corresponds to the chemical potential $\mu$, which is a step function \cite{Sagert:2005fw} at vanishing temperature. The number density of electrons $n_e$ is 
\beq
n_e=\int_0^{k_F}\frac{2}{(2 \pi \hbar )^3}d^3k \quad \rightarrow \quad n_e=\frac{k_F^3}{3 \pi^2 \hbar^3}
\eeq
with $k_F$ as the Fermi momentum. The mass density $\rho=n_e m_N \cdot \frac{A}{Z}$ with $m_N$ as the nucleon mass and $\frac{A}{Z}=2$ for a $^{12}C$ White Dwarf star. From these assumptions it follows that 
\beq
k_F=\hbar\sqrt[3]{\frac{3\pi^2\rho Z}{m_N A}}.
\eeq
The electrons in a white dwarf are mostly responsible for the degeneracy pressure whereas the nucleons contribute to the star's mass.
The energy density is 
\beq
\epsilon=\frac{8 \pi }{(2\pi\hbar)^3}\int_0^{k_F}E(k)k^2 dk
\eeq
where $E(k)=\sqrt{k^2c^2+m^2c^4}$ from the relativistic energy momentum relation. The derivation of the corresponding pressure integral can be found in \cite{Sagert:2005fw} and reads 
\beq
p=\frac{8 \pi }{3(2\pi\hbar)^3}\int_0^{k_F} \frac{k^2c^2}{E(k)}k^2 dk.
\eeq
Solving the integrals and following the procedure in \cite{Silbar04,Sagert:2005fw} one arrives at a polytropic EoS of the form 
\beq \label{polytropic_eos}
p=K\epsilon^{\Gamma}
\eeq
with $K=\rm{const.}$ for an adiabatic index $\Gamma$. The adiabatic index is commonly given as $\Gamma=1+1/n$, with n as the polytropic index. 
K can be determined via the relation 
\beq \label{Katznklo}
K=p_0 \cdot \epsilon_0^{-\Gamma}
\eeq 
with $p_0=1.6022\cdot 10^{32}~\rm{Pa}$
which is equal to $1~\rm{MeV/fm^3}$ or $1.3234 \times 10^{-6}~\rm{km^{-2}}$, depending on the units one wants to use. The energy density may be chosen to $\epsilon_0=94.38~\rm{MeV/fm^3}$ which corresponds to a baryon density of $n_0=0.1~\rm{fm^{-3}}$, see also \cite{Glendenning_book,Postnikov:2010yn}.\\
For a given $\Gamma$ the value of K determines the maximum mass in the mass radius relation. 
It turns out that mass and radius scale with the constant K to some power. The relation of $\rm{M/R}$ is called compactness $C$. In Sec. \ref{landsackschlack} we show that C is independent on K.
\subsubsection{The Lane-Emden Equation}\label{Lahme_Ente}
The Lane Emden equation is a dimensionless Poisson equation for a spherically symmetric mass distribution. This mass distribution can be described via a polytrope for example, see eq.~(\ref{polytropic_eos}). A dimensionless treatment enables us to study scaling solutions of the corresponding mass radius relations, i.e. when solving for eqs.~(\ref{mass_appeal_madness}) and (\ref{hydro_schraubenschluessel}). We will encounter scaling solutions again in Sec. \ref{landsackschlack}.\\
Rearranging eq.~(\ref{hydro_schraubenschluessel}) and deriving with respect to $r$ yields
\begin{equation}\label{first_step_towards_wurst}
 \frac{d}{dr}\left[\frac{r^2}{\rho(r)}\frac{dP}{dr}\right]=-\frac{Gdm(r)}{dr}, \qquad {\rm{using} \, eq.~(\ref{mass_appeal_madness})} \qquad \frac{1}{r^2}\frac{d}{dr}\left[\frac{r^2}{\rho(r)}\frac{dP}{dr}\right]=-4 \pi G \rho(r).
\end{equation}
Making use of the polytropic EoS eq.~(\ref{polytropic_eos}) and its derivative $dp/dr$ to substitute these quantities in eq.~(\ref{first_step_towards_wurst}), one arrives at 
\begin{equation}\label{first_step_towards_wurstalikerscher}
 \frac{\Gamma K}{r^2}\frac{d}{dr}\left[r^2\rho(r)^{\Gamma-2}\frac{d\rho(r)}{dr}\right]=-4 \pi G \rho(r).
\end{equation}
In order to study scaling solutions, the equation above needs to become dimensionless. To this purpose one introduces
$\rho=\rho_c \theta^n$ as the density in units of the central density $\rho_c$ and $0\leq \theta^n \leq 1$ as the corresponding function depending on the radius $r$. The radius itself becomes $r=\xi \lambda_n$ where $\lambda_n$ is in units of a characteristic lenghth. $\xi$ is the dimensionless radius. Plugging these values in eq.~(\ref{first_step_towards_wurstalikerscher}) one finds
\begin{equation}\label{pippilangstomp}
\underbrace{(n+1)\frac{K \rho_c^{\frac{1-n}{n}}}{4 \pi G}}_{\rm{lenghth^2}} 
\cdot \underbrace{\frac{1}{r^2}}_{\rm{\frac{1}{lenghth^2}}}
\cdot \underbrace{\frac{d}{dr}\left[ r^2\frac{d\theta}{dr}\right]}_{\rm{dimensionless}}=
\underbrace{-\theta^n}_{\rm{dimensionless}}.
\end{equation}
Due to dimensional reasoning the characteristic lenghth is
\begin{equation}
 \lambda_n=\left((n+1)\frac{K \rho_c^{\frac{1-n}{n}}}{4 \pi G}\right)^{1/2}.
\end{equation}
Eventually one arrives at the Lane Emden equation
\begin{equation}\label{emdenerstrasse9}
 \frac{1}{\xi^2}\frac{d}{d \xi}\left(\xi^2\frac{d \theta}{d \xi} \right)=-\theta^n
\end{equation}
with $\theta$ a dimensionless density and $\xi$ a dimensionless radius. With the Lane Emden equation one is able to find mass radius relations  adopting eq.~(\ref{mass_appeal_madness}). The high density limit corresponds to $\Gamma=4/3$ and yields the famous Chandrasekhar mass limit \cite{Sagert:2005fw} for White dwarf stars with $M_{WD}\lesssim 1.4~M_{\odot}$ independent on the central pressure or the radius of the star. 
As the density in the White dwarf increases, the electrons become more relativistic until the mass limit is reached. The Chandrasekhar mass limit represents the maximum possible mass for a White dwarf for a purely relativistic polytropic EoS.\\
There are three known analytical solutions ($n=0$, $n=1$ and $n=5$) for the Lane Emden equation which are solutions of the cylindrical Bessel functions. We will further investigate analytical solutions for $n=0$ and $n=1$ in Sec.~\ref{karmageddon}.
\subsection{The EoS for selfbound stars}\label{mit_bag}
A selfbound star is in no need for gravity to be stable. It stabilizes itself due to 
attractive interactions, similar to how a nucleus does not need gravity to be stable \cite{Glendenning:1984jr,Gilson93}. For the EoS this  implies that for zero pressure the value of the energy density is nonzero. The corresponding mass radius relation scales as $M \propto R^3$.\\ 
Selfbound stars may be realized in nature as pure quark stars \cite{Ivanenko:1965dg,Itoh:1970uw,Haensel:1986qb,Olinto87,Bodmer:1971we,Witten:1984rs,Schertler:1999xn,Schertler:2000xq,Zacchi:2015oma,Zacchi:2016tjw,Zacchi:2019ayh} and are still not ruled out yet due to observation or any astrophysical constraint known so far.
Though there are several sophisticated ways
to implement effects of confinement by means of quantum field theory, for our purpose it is sufficient to use a pure phenomenological model: The MIT-bag model \cite{Chodos74}.
The MIT-bag model is an intuitive quark model and has already been developed in the 70’s by physicists
of the Massachusetts Institute of Technology to describe the structure of hadrons. The central idea for
modelling confinement is that quarks with flavour f are contained inside a finite volume in vacuum. Inside this \textit{Bag} they behave as a free Fermi gas. The \textit{Bag} carries the quarks and is colour neutral from the outside.  
The so called Bag constant B models the pressure from the vacuum onto the bag. The 
quarks have to compensate the pressure resulting of B by their kinematic pressure.
This translates into the relation
\begin{eqnarray}
 p=\sum_f p_f-B \qquad{\rm{and}} \qquad \epsilon= \sum_{f} \epsilon_f+B.
\end{eqnarray}
For a relativistic massless gas of particles $\epsilon_f=3p_f$ \cite{Sagert:2005fw,Weissenborn:2011qu}. Expressing the pressure $p$ in terms of the energy density $\epsilon$ gives
\beq \label{selfdestruction}
p(\epsilon)=\frac{1}{3}(\epsilon-4B).
\eeq
We will encounter this EoS and the corresponding compact stars and their properties in the following chapters. Note that this EoS is not being used for any classical analysis. 
It is the general solution for an ultrarelativistic gas and hence of importance as a limiting case in general relativity \cite{Sagert:2005fw}.
\section{Classical Quadrupole deformation}\label{quadrupole}
The trajectory of a planet around a star such as the Sun is an ellipse. The influence of other planets is responsible for a small deviation of the regular ellipse trajectory. It was found that the planets obey the laws found by Kepler, but the perihelion shift of mercury was a mystery. The perihelion shift was determined already in 1859 by LeVerrier to be 5.74 arc.sec per year. With Newtonian physics the value should have been 5.31 arc.sec per year, and the deviation of 0.43 arc.sec per year could not be explained.\\
Before the discovery of general relativity, which gives the solution to this riddle, it was speculated that the deformation of the sun due to rotation is responisble for this effect, at least to some extent. It turned out that the deformation due to the so called quadrupole moment is responsible only for less than a percent of the perihelion shift of mercury.\\ 
A spherically symmetric mass distribution may, without loss of generality, be bulged around the equator. Then of course the mass distribution is no longer spherically symmetric, see also fig.~\ref{ellipse}. The important symmetry left is the rotation around the z-axis (in the x- and y direction only rotations $\propto 2 k \pi$ are allowed). Unless as in electrodynamics, a dipole moment does not exist, but such a mass distribution has a quadrupole moment.
Higher order momenta obey other symmetries which we neglect.  
\subsection{Deformation due to rotation}\label{thangorodrim}
\begin{figure}[H]
\begin{center}
\includegraphics[width=.5\columnwidth]{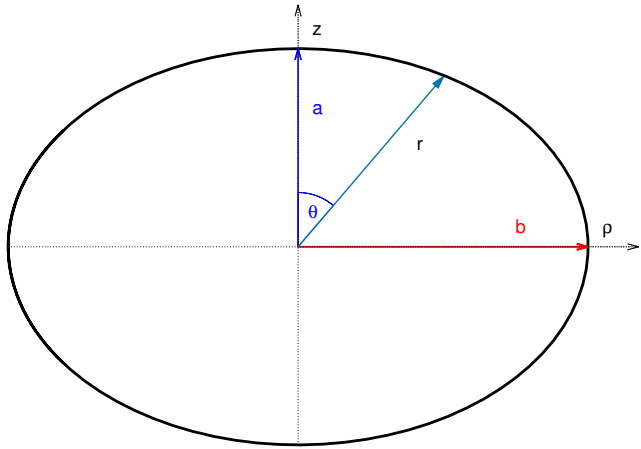}
\caption{\textit{A cut through an ellipsoid gives an ellipse with the quantities $a=R\left(1-\frac{2}{3}E\right)$ on the $\rho$-axis, $b=R\left(1+\frac{1}{3}E\right)$ on the z-axis and the enclosed angle $\theta$ according to eqs. \ref{ella} and \ref{machtnesperre}. Rotation around the z-axis yields a dense bulge at the equator and hence a quadrupole moment.}}
\label{ellipse}
\end{center}
\end{figure}

For an ordinary M-star such as our sun, which needs roughly one month to rotate around its axis, the deviation due to rotation of the radius $r$ from a spherically symmetric sun with radius $r$ is rather small, but can be estimated nonetheless. 
An ordinary star surface can be described via 
\begin{equation}
\vec{r}(\theta)=\left(\begin{array}{c}
r(\theta) \sin \theta\\
r (\theta) \cos \theta
\end{array}\right),
\end{equation}
where $\theta$ is the angle between the northpole and the equatorial plane. The coordinates used are those when cutting through a sphere. The surface is described then by the corresponding circle with the coordinates $\rho$ and z arranged in a vector, see also fig.~\ref{ellipse}.
The radial change is the derivative with respect to $\theta$
\begin{equation}
\frac{d{\vec{r}}(\theta)}{d \theta}=r'\left(\begin{array}{c}
\sin \theta\\
\cos \theta
\end{array}\right) + 
r\left(\begin{array}{c}
\cos \theta\\
-\sin \theta
\end{array}\right).
\end{equation}
The centrifugal acceleration $\omega^2 r$ perpendicular to the surface of course also depends on $\theta$, so that  
the total acceleration at the surface of the sun is 
\begin{equation}
\vec{F}=\left(\begin{array}{c}
-\omega^2 r \sin \theta - g\sin \theta\\
-g\cos \theta
\end{array}\right) \propto \vec{n},
\end{equation}
where $\vec{n}$ is perpendicular to the surface $r(\theta)$ and $g\simeq 270~\rm{m/s^2}$ is the gravitational acceleration at the  surface of the sun. 
The change of $r$ due to rotation is $d\vec{r}(\theta)/d \theta$. 
Due to perpendicularity $\vec{n} \cdot d\vec{r}(\theta)/d \theta=0$, so that
\begin{eqnarray}
 (\omega^2 r \sin \theta - g \sin \theta)({r'}\sin \theta+r\cos\theta)&=&g\cos\theta({{r}'}\cos\theta-r \sin\theta)\\
 r'=\frac{\omega^2 r}{g}\sin \theta \frac{d}{d \theta}(r\sin\theta)\bigg \vert \int &\Rightarrow& 
 rg-\frac{\omega^2}{2}\left(r\sin \theta \right)^2=C=:\tilde{R}g\\
\end{eqnarray}
The integration constant $C$ can be determined to be 
\begin{eqnarray}  \label{luisvulture}
C/g ={r(\theta=0)}=r-\tilde{R}  &=& \frac{\omega^2 r^2}{2g}\sin^2\theta
= \frac{\omega^2 r^2}{2g}(1-\cos^2 \theta) 
\end{eqnarray} 
The ellipse equation taken from A.E.H. Love \cite{Love:1908tua}, which
is given here for convenience, is 
\begin{equation}\label{sportboat}
 r=R\left[1+E\left(\frac{1}{3}-\cos^2\theta\right) \right],
\end{equation}
and further evaluation at the north pole, where $\theta=0$, and at the equator, where $\theta=\pi/2$, yields  
\begin{eqnarray}\label{ella}
\theta&=&0 \Rightarrow r=R\left(1-\frac{2E}{3}\right)\\ \label{machtnesperre}
\theta&=&\frac{\pi}{2} \Rightarrow r=R\left(1+\frac{E}{3}\right).
\end{eqnarray}
These results tell us that the oblateness (ellipticity) E of an ellipse due to rotation is 
\begin{equation}\label{centri_cani_gatti}
 E=\frac{\omega^2R}{2g}
\end{equation}
\subsection{Deformation due to an external field}
We begin the calculation with an acceleration similar as in section \ref{thangorodrim}, only now we consider an external field causing the deformation of the star. The effect of an external field due to a quadrupole deformation is described via the quadrupole tensor $Q_{ij} \propto \epsilon_{ij}$, see eq.~(\ref{ruchnefisch}). For simplicity we assume only $\epsilon_{33}\neq 0$ which models the rotation around the z axis. For
\begin{equation}
\vec{F}=\left[-\left(\begin{array}{c}
0\\
z
\end{array}\right)\epsilon_{33} - g \left(\begin{array}{c}
\sin\theta\\
\cos\theta
\end{array}\right)\right]\propto \vec{n},
\end{equation}
which is again proportional to $\vec{n}$. It is hence anew valid to say that $\vec{n} \cdot d\vec{r}(\theta)/d \theta=0$. Performing an analoguous calculation as in the previous section \ref{thangorodrim}, one arrives at
\begin{equation}
 r-\tilde{R}=-\frac{\epsilon_{33}}{2g}r^2\cos^2\theta.
\end{equation}
Again comparing with eq.~(\ref{sportboat}) for two different angles 
\begin{eqnarray}
\theta=0 \Rightarrow r=R\left(1-\frac{2}{3}E\right) \qquad {\rm{and}} \qquad
\theta=\frac{\pi}{2} \Rightarrow r=R\left(1+\frac{1}{3}E\right)
\end{eqnarray}
gives
\begin{equation}\label{mrtwurst}
 E=\frac{\epsilon_{33}R}{2g}
\end{equation}
as the excentricity E of an ellipse, i.e. the bulging, due to an external field.\\ 
The main statement is that rotation corresponds qualitatively to an external gravitating field $\omega^2 \propto \epsilon_{33}$ caused by a companion star. This is seen when comparing eq.~(\ref{centri_cani_gatti}) with eq.~(\ref{mrtwurst}).
\subsection{Quadrupole moments}
We are now ready to take a closer look at the quadrupole moments. 
The starting point to determine the quadrupole moments is the 
traceless quadrupole tensor in cartesian coordinates
\begin{equation}\label{tracelessmofo}
 Q_{ij}=\int d\varphi \int dr r^2 \int d\theta \sin(\theta) [3x_{j}x_{j}-r^2\delta_{ij}]\rho(\theta)
\end{equation}
where $\rho(\theta)$ is the corresponding density at a given angle. Because of symmetry reasons the calculation is independent on $\varphi$, so that $Q_{ij}=0$ for $i \neq j$ and $Q_{11}=Q_{22}$. 
Furthermore $Q_{33}=-2Q_{11}=-2Q_{22}$, which is not surprising since $Q_{ij}$ is traceless.\\  
To simplify the integration, we first use a constant density $\rho_0$ throughout the calculation 
\begin{eqnarray}
 Q_{33}&=&\int_0^{2\pi} d\varphi \int_0^{\pi} d\theta \sin(\theta) \int_0^{R(\theta)} dr r^2 \cdot r^2 [3\cos(\theta)-1]\rho_0\\
 &=&\frac{2\pi\rho_0}{5}\int_0^{\pi} d\theta\sin\theta\left(3\cos^2\theta-1\right)R(\theta)^5.
\end{eqnarray}
$R(\theta)^5$ is approximately given as $R(\theta)^5\simeq R^5\left[1+5E\left(\frac{1}{3}-\cos^2\theta\right) \right]$ from the definition given in eq.~(\ref{sportboat}), and $z=r\cos\theta$ so that 
\begin{eqnarray}
 Q_{33}&=&\frac{4\pi \rho_0}{5} \int_0^1 dz (3z^2-1)R^5\left[1+5E\left(\frac{1}{3}-z^2\right) \right].     
\end{eqnarray}
Solving the integral for constant density $\rho_0$ yields 
\begin{eqnarray}\label{quaddrruuppohl}
 Q_{33}&=&-\frac{4}{5}M R^2 E.     
\end{eqnarray}
The quadrupole term in the gravitational potential is
\begin{equation}\label{kampen}
 \Phi_2/G=\frac{x_{i}Q_{ij}x_{j}}{2r^5}=\frac{1}{2r^5}\left((x^2+y^2)Q_{11}+z^2Q_{33}\right)=\frac{Q_{33}}{4r^3}\left( 3\cos^2\theta-1\right).
\end{equation}
The full potential in multipole expansion $\Phi=\Phi_0+\Phi_2+ \dots$ reads 
\begin{equation}\label{kampen2}
 \Phi=-\frac{GM}{R}+J_2\frac{GM}{R}\left[\frac{R}{r}\right]^2 \frac{3\cos^2\theta-1}{2}+\dots
\end{equation}
where $J_2$ is the quadrupole moment parameter for $\Phi_2$
\begin{equation}\label{quadriceps_beast}
 J_2=\frac{Q_{33}}{2MR^2}=-\frac{2}{5}E.
\end{equation}
It turns out that $J_2$ is independent on the density. It only depends on the radius, the surface gravity 
and in case of rotation, on the angular velocity. In case of an external tidal field $J_2 \propto \epsilon_{33}$ instead of $\omega^2$. 
Inserting the values for our sun in equation (\ref{centri_cani_gatti}), one finds that $J_2\simeq 3.7 \cdot 10^{-6}$.
This roughly determined value is in good accordance with the oblateness parameter $C_2$ found in \cite{Kuhn1638}, and matches the values discussed in \cite{2001SoPh199217G}, deduced mainly in the seventies and eighties, (even for constant density) to a good approximation. The value of $J_2$ by all means depends on the utilized model and is also dependent on the mass- and density distribution of the star, which we neglected in our first approach. Moreover, such a small value of the quadrupole moment parameter cannot be responsible for the perihelion shift of mercury.\\
Eq.~(\ref{quaddrruuppohl}) represents the quadrupole deformation due to rotation, which is $\propto \omega^2$ or due to the appearence of an external tidal field which is $\propto \epsilon_{33}$. More generally
\beq\label{ruchnefisch}
Q_{ij}=-\lambda \epsilon_{ij},
\eeq
with the parameter $\lambda$ as the tidal polarizability, i.e. the ratio
of the induced quadrupole $Q_{ij}$ to the perturbing
tidal field $\epsilon_{ij}$ from the companion star. The relation of $\lambda$ to the Love number $k_2$ according to \cite{Hinderer:2007mb,Flanagan:2007ix,Hinderer:2009ca,Postnikov:2010yn} and Refs. therein is
\beq\label{ruchnefischali}
\lambda =\frac{2R^5}{3G}k_2
\eeq
with $R$ the radius of the star and $G$ the gravitational constant.
For an relativistic approach, the tidal deformability parameter $\Lambda$ depends on the compactness $C$ of the compact star and on the Love number $k_2$ \cite{Love:1908tua,Hinderer:2007mb,Postnikov:2010yn} via 
\begin{equation}\label{loveandloveismierdadelburro}
 \Lambda=\frac{2 k_2}{3 C^5}.
\end{equation}
Classically $C=0$ and the tidal deformability is
described only by the Love number $k_2$.
However, with a constant surface gravity $g=GM/R^2$ and eq.~(\ref{mrtwurst}), $Q_{33}$ from eq.~(\ref{ruchnefisch}) becomes
\beq\label{wiegehtdenndas}
Q_{33} = -\frac{2R^5}{5G}\epsilon_{33}.
\eeq
We have eventually found a connection of the eccentricity of a deformed star due to an induced external field, most commonly induced by companion star, with the tidal polarizability $\lambda$ or the Love number $k_2$. For a constant density  
\beq\label{fleischgewordenerhass}
\lambda=\frac{2R^5}{5G} \qquad {\rm{and}} \qquad k_2=\frac{3}{5}.
\eeq
Classically, the Love number is a constant and does not depend on any other quantity for constant density \cite{Love:1908tua}. In the original work and in the notation from A.E.H. Love from 1906 \cite{Love:1908tua} he finds 
a value of $H(a=r)=h \simeq 3/5$ as a solution of his analysis concerning the sun, earth and moon system.\\
He interprets his results as follows:\\ 
\textit{The inequality produced in
the potential of the Earth near its surface by the action of the Sun and Moon
is about $k=4/15$ of the tide-generating potential, and the inequality
produced in the surface of the Earth according to the relation $h-k=1/3$ is about $h=3/5$ of the true
equilibrium height of the tide. 
If the matter within the Earth is assumed to be absolutely incompressible and of uniform density $\rho$, one should have corrections due to the rigidity of matter} (quoted from Ref.~\cite{Love:1908tua}).\\ 
This is indeed the case, as we will see.
However, the quadrupole moment parameter for constant density can also be expressed as
\beq
J_2=\frac{-\lambda \epsilon_{33}}{2MR^2}=\frac{R^3\epsilon_{33}}{5 G M}.
\eeq
Love numbers have also been determined for different celestial bodies in our solar system.
An inner planet experiences contributions that arise from the tidal- and rotational 
bulges of the surrounding objects. Precession can in principle be neglected \cite{Ragozzine:2008bp}, but leading effects from general relativity cannot be disregarded \cite{Jordan:2008zi}. 
Moons or satellites are also being considered for the determination of $k_2$, which is of course a nontrivial business because 
the extraction of the Love number $k_2$ might be further complicated due to oceans under the crust and hence the layer structure, i.e. which material is present at which depth of the celestial body. 
The value of the Love number for an ``elastic'' Earth is about $k_2\simeq 0.3$. 
The Love number of our Moon can be modelled by a mixture of a fluid 
and a solid core and is about ten times smaller \cite{Lainey_2016}.
Within our solar system, 
Titan has a particularly large Love number. This feature implies that Titan
is highly deformable which is consistent with a global ocean under the ice
crust of Titan \cite{article_titan}. Derivations and further discussion on that subject can be found in Refs.~\cite{10.1093/mnras/99.5.451,Wu:2001wn,article_titan,Ragozzine:2008bp,Lainey_2016}.


\subsection{Hydrostatic equilibrium}\label{fressoderwatt}
\begin{figure}[H]
\begin{center}
\includegraphics[width=.6\columnwidth]{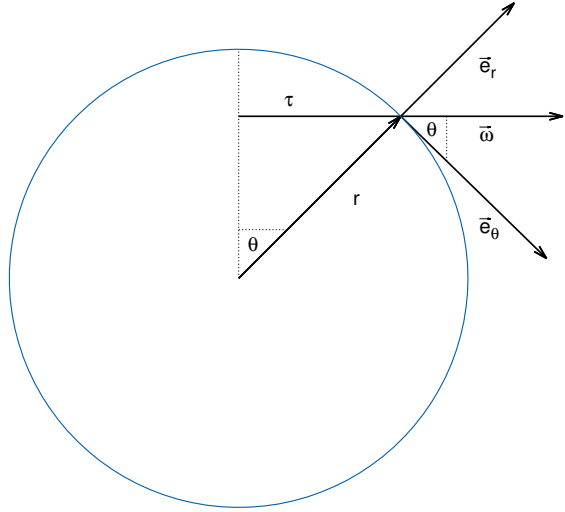}
\caption{\textit{The construction of $\vec{\omega}$ with the corresponding unit vectors $\vec{e}_{\theta}$ and $\vec{e}_{r}$. $\vec{e}_{\phi}$ is perpendicular to $\vec{\omega}$ pointing inwards or outwards the plane. $\tau$ is the distance from a point at the surface to the rotational axis.}}
\label{kugelflaechenfucktionen}
\end{center}
\end{figure}

Let us have a look now how to determine appropriate equations which describe such tidal deformations. 
We follow the procedure in Refs.~\cite{roxburgh,Paterno:1996abc,2001SoPh199217G} 
to derive the quadrupole related expressions classically. 
Starting point is the pressure gradient
\beq \label{gradient_pressure}
\nabla P=\rho(r) \left[-\nabla \Phi(r) + \Omega^2\vec{\omega}  \right],
\eeq
where $\rho(r)$ is the matter density at distance $r$, $\Phi(r)=-GM/r$ the gravitational potential, $G$ the gravitational constant, $M$ the mass, $\Omega$ the rotational frequency and $\vec{\omega}$ the rotational vector pointing perpendicular outwards of the spheres surface.
$\vec{\omega}$ needs to be constructed through the unit vectors, so that $\vec{\omega}=\tau \hat{e}_{\tau}=r\sin(\theta)\hat{e}_{\tau}$, $\tau$ being the distance to the rotational axis, see figure \ref{kugelflaechenfucktionen}. $\hat{e}_{\tau}$ is 
\beq
\hat{e}_{\tau}=\hat{e}_{\omega}=\sin(\theta)\hat{e}_{r}+\cos(\theta)\hat{e}_{\theta},
\eeq
For $\theta=0$ the $\hat{e}_r$ contribution vanishes since the centrifugal force does not elevate a body placed on top of the sphere, when rotating around the z axis.
The term representing the centrifugal force in eq.~(\ref{gradient_pressure}) can thereby be rewritten as
\beq \label{omega_substitute}
\Omega^2\vec{\omega}=\Omega^2 r \sin(\theta)\left[ \sin(\theta)\hat{e}_{r}+\cos(\theta)\hat{e}_{\theta} \right].
\eeq
Inserting eq.~(\ref{omega_substitute}) in eq.~(\ref{gradient_pressure}) and
rewriting everything in spherical coordinates to compute the gradient in eq.~(\ref{gradient_pressure}), one arrives at
\begin{eqnarray}
 \frac{\partial P}{\partial r}\hat{e}_{r}+ \frac{1}{r}\frac{\partial P}{\partial \theta}\hat{e}_{\theta}&=& \rho\left[ -\frac{\partial \Phi}{\partial r}\hat{e}_{r} -\frac{1}{r}\frac{\partial \Phi}{\partial \theta}\hat{e}_{\theta}+ \Omega^2 r \sin(\theta)\left( \sin(\theta)\hat{e}_{r}+\cos(\theta)\hat{e}_{\theta} \right) \right].
\end{eqnarray}
Eventually
\begin{eqnarray} \label{paterno_1}
 \frac{\partial P}{\partial r}&=&\rho \left[-\frac{\partial \Phi}{\partial r} + \Omega^2 r \sin^2(\theta)\right]\\ \label{paterno_2}
 \frac{\partial P}{\partial \theta}&=&\rho \left[-\frac{\partial \Phi}{\partial \theta} + \Omega^2 r^2 \sin(\theta)\cos(\theta)\right].
\end{eqnarray}
At this point we make use of the Legendre Polynomials. Here $P_0(x)=1$ for a perfect sphere, $P_1(x)=x$, but in our case $P_1(x)=0$ since we have no dipole and eventually $P_2(x)=\frac{1}{2}(3x^2-1)$ is a small deviation from a perfect sphere giving
\bea \label{pzweiohnestrich}
P_2(\theta)&=&=\frac{1}{2}\left(3 \cos^2(\theta)-1 \right)=\frac{1}{2}\left(2-3\sin^2(\theta)\right) \\ \label{pzweistrich} 
P_2'(\theta)&=&-3\sin(\theta)\cos(\theta) \\ \label{pzweizweistrich}
P_2''(\theta)&=&-3\left(2 \cos^2(\theta)-1\right).
\eea
Rearranging eq.~(\ref{pzweiohnestrich}) and plugging eqs.~(\ref{pzweiohnestrich}) and (\ref{pzweistrich}) in eqs.~(\ref{paterno_1}) and (\ref{paterno_2}) we remain with
\bea \label{paterno_alienation}
\frac{\partial P}{\partial r} &=& \rho \left[ -\frac{\partial \Phi}{\partial r} + \frac{2 \Omega^2 r}{3} \left(1-P_2(\theta) \right) \right]\\ \label{paterno_alienation2}
 \frac{\partial P}{\partial \theta} &=& \rho \left[ -\frac{\partial \Phi}{\partial \theta} - \frac{\Omega^2 r^2}{3} P'_2(\theta) \right].
\eea
These equations, eq.~(\ref{paterno_alienation}) and (\ref{paterno_alienation2}), can also be found in Ref.~\cite{Paterno:1996abc}, there as eq.~(5) and eq.~(6).
For continuous functions one can cross differentiate eqs.~(\ref{paterno_1}) and~(\ref{paterno_2}), so that
\begin{eqnarray} \label{paterno_3}
 \frac{\partial}{\partial \theta} \frac{\partial P}{\partial r}&=&\frac{\partial}{\partial \theta}\left(\rho \left[ -\frac{\partial \Phi}{\partial r} + \frac{2 \Omega^2 r}{3} \left(1-P_2(\theta) \right) \right]\right)\\ \label{paterno_4}
 \frac{\partial}{\partial r} \frac{\partial P}{\partial \theta}&=& \frac{\partial}{\partial r}\left( \rho \left[ -\frac{\partial \Phi}{\partial \theta} - \frac{\Omega^2 r^2}{3} P'_2(\theta) \right]\right).
\end{eqnarray}
Now one can equalize eqs.~(\ref{paterno_3}) and~(\ref{paterno_4}). We have to bear in mind that $\rho$ depends on either angle and radius, i.e. $\rho(r, \theta)$.\\
After a bit of non-enlightening algebra 
\bea \label{wassollderscheiss}
- \frac{\partial \rho}{\partial r} \frac{\partial \Phi}{\partial \theta}+\frac{\partial \rho}{\partial \theta} \frac{\partial \Phi}{\partial r}=\frac{2 \Omega^2 r}{3} \frac{\partial \rho}{\partial \theta}\left[ 1 - \rho P_2(\theta)\right] + \frac{1}{3}\Omega^2 r^2 \frac{\partial \rho}{\partial r} P_2'(\theta).
\eea
Following ref.~\cite{Paterno:1996abc} we expand 
\bea \label{colazero}
\Phi(r,\theta)&=&\Phi_0(r)+\Phi_2(r)P_2(\theta) \qquad {\rm{and}} \\ \label{firstcomefirstserve}
\rho(r,\theta)&=&\rho_0(r)+\rho_2(r)P_2(\theta).
\eea
Plugging everything in and rearranging eq.~(\ref{wassollderscheiss}) yields
\beq \label{morgoth_bauglir}
\rho_2(r) \frac{\partial \Phi_0}{\partial r}=\Phi_2 \frac{\partial \rho_0}{\partial r} + \frac{1}{3} \Omega_0^2 r^2\frac{\partial}{\partial r}\left(\rho_0 \right)
\eeq
which is eq.(7) in Ref.~\cite{Paterno:1996abc}. Quadratic terms with index 2 such as $\rho_2(r)P_2'(\theta)$ are neglected since these terms are $\ll 1$.
\subsection{Poisson equation}
The Poisson equation in Newtonian physics is the source term of gravity
\beq \label{source_of_wurst}
\nabla^2 \Phi=4 \pi \rho G. 
\eeq
where we expand the potential $\Phi(r, \theta)$ and the density $\rho(r, \theta)$ according to eqs.~(\ref{colazero}) and (\ref{firstcomefirstserve}), see also Refs.~\cite{Paterno:1996abc,2001SoPh199217G}. Using the second derivatives in spherical coordinates
\beq
\nabla^2 \Phi=\frac{2}{r}\Phi_0'(r)+\Phi_0''(r)+\Phi_2''(r)P_2(\theta)+\frac{2}{r}\Phi_2'(r)P_2(\theta)+\frac{1}{r^2 \tan(\theta)}\Phi_2(r)P_2'(\theta)+\frac{1}{r^2}\Phi_2(r)P_2''(\theta)=
4 \pi \left[\rho_0 +\rho_2(r)P_2(\theta)\right] G. 
\eeq
When substracting the solution for a homogenous and perfect sphere only the perturbation remains, so that
\beq \label{langesteilamkopf}
P_2(\theta)\left[ \Phi_2''(r)+\frac{2}{r}\Phi_2'(r) - 4 \pi G \rho_2(r)\right] + \frac{1}{r^2 \tan(\theta)}\Phi_2(r)P_2'(\theta)+\frac{1}{r^2}\Phi_2(r)P_2''(\theta)=0.
\eeq
Expressing eq.~(\ref{pzweizweistrich}) in terms of eq.~(\ref{pzweiohnestrich}) gives
\beq \label{fingolfin}
P_2''(\theta)=-4P_2(\theta)+1.
\eeq
The eqs.~(\ref{pzweistrich}), (\ref{fingolfin}) and the cosine squared from eq.~(\ref{pzweiohnestrich}) enter in eq.~(\ref{langesteilamkopf}). Rearranging for $\rho_2(\theta)$ gives
\beq \label{melkor}
\rho_2(r)=\frac{1}{4 \pi G}\left[\Phi_2''(r)+\frac{2}{r}\Phi_2'(r)-\frac{6 \Phi_2(r)}{r^2} \right].
\eeq
Now $\rho_2(r)$ from eq.~(\ref{morgoth_bauglir}) enters in eq.~(\ref{melkor}). At this point one needs to recall that 
$\Phi(r)=\Phi_0(r)=-GM/r$. Rearranging and equalizing yields
\beq \label{dichtegleichung}
\rho_2(r)=\left[ \Phi_2 \frac{\partial \rho_0}{\partial r} + \frac{1}{3} \Omega_0^2 r^2\frac{\partial}{\partial r}\left(\rho_0 \right) \right]\cdot \frac{r^2}{G M}=\frac{1}{4 \pi G}\left[\Phi_2''(r)+\frac{2}{r}\Phi_2'(r)-\frac{6 \Phi_2(r)}{r^2} \right].
\eeq
With the substitutions 
\bea \label{substituete_1}
x=r/R_{\odot}, \quad U(x)=\frac{4\pi r^3 \rho_0}{M_r}, \quad M_r=4 \pi \int_0^r \rho_0r'^2dr', \\ \label{substituete_2}
V(x)=\frac{d \ln(\rho_0)}{d \ln(r)}, \quad \omega(x)=\frac{\Omega_0(r)}{\Omega_{*}} \quad {\rm{and}} \quad y(x)=\frac{\Phi_2(r)}{\Omega_{*}^2 R_{\odot}}, 
\eea
where $\Omega_{*}$ is a reference angular velocity, the quantities in eq.~(\ref{dichtegleichung}) become dimensionless. Remember that dimensionless quantities enable us to study scaling solutions (see Sec.~\ref{Lahme_Ente}: The Lane-Emden Equation).
\subsubsection{Case: No rotation}
In case one neglects rotation $\Omega_0=0$, so that eq.~(\ref{dichtegleichung}) simplifies to 
\beq \label{dichtegleichung_2}
\Phi_2 \frac{\partial \rho_0}{\partial r} \frac{4 \pi r^2}{M}=\left[\Phi_2''(r)+\frac{2}{r}\Phi_2'(r)-\frac{6 \Phi_2(r)}{r^2} \right].
\eeq
It follows that 
\beq \label{hillary_clinton}
\Phi_2''(r)+\frac{2}{r}\Phi_2'(r)-\left[ 6+U(r)V(r)\right]\frac{\Phi_2(r)}{r^2}=0.
\eeq
We will have a closer look at eq.~(\ref{hillary_clinton}) in the next section~\ref{bobgeldorf}.
\subsubsection{Case: Nonzero Rotation}
If now $\Omega_0 \neq 0$ the calculation is slighlty longer. One intermediate step is
\beq
\frac{1}{4 \pi G}\Omega_{*}^2\left[\frac{1}{x^2}\frac{d}{dx}\left(x^2 \frac{dy}{dx}\right) - \frac{6y}{x^2}  \right]=\frac{1}{d \Phi_0/dx}\left[\Omega_{*}^2 R_{\odot}^2 y \frac{d \rho_0}{dx}+\frac{1}{3}x^2R_{\odot}^2\frac{d}{dx}\left(\rho_0\Omega_0^2 \right) \right],
\eeq
with $d \Phi_0/dx=G M_r R_{\odot}/r^2$, $\rho_0=U(x) M_r/4 \pi r^3$ and $d \rho_0/dx=\rho_0 V(x)/ x$. The result is finally 
\beq \label{bryan_adams}
\frac{1}{x^2}\frac{d}{dx}\left( x^2\frac{dy}{dx}\right)-\left[ 6+U(x)V(x)\right] \frac{y}{x^2}=\frac{U(x)}{3}\left( V(x)\omega^2+2x\omega \frac{d \omega}{dx} \right),
\eeq
which is eq.~(9) in Ref.~\cite{Paterno:1996abc} and which has already been derived in 1964 by Roxburgh \cite{roxburgh}. Eq.~(\ref{bryan_adams}) can only be solved numerically. In case of no rotation $\omega=0$ and eq.~(\ref{bryan_adams}) reduces to eq.~(\ref{hillary_clinton}).
One may recognize the familiarity with the Lane Emden equation, eq.~(\ref{emdenerstrasse9}). At this point it may be interesting to note that many problems in astrophysics (and physics) generally reduce to differential equations which are closely related to solutions of Bessel functions, or solutions of Bessel functions itself.
\subsection{Love number and tidal deformations}\label{bobgeldorf}
Let us now have a closer look at eq.~(\ref{hillary_clinton}) and solve 
the differential equation first of all for $UV=0$. Hence 
\begin{equation}\label{diff_eq_del_cazz}
\Phi_2''(r)+\frac{2}{r}\Phi_2'(r)-\frac{6}{r^2}\Phi_2(r)=0. 
\end{equation}
Equation~(\ref{diff_eq_del_cazz}) can be solved via the ansatz $H(r)=\alpha r^n$, so that
\begin{eqnarray}\label{diff_eq_del_cazz_sol}
n(n+1)\alpha r^{n-2}+2n\alpha r^{n-2}-6\alpha r^{n-2}&=&0 \qquad \Rightarrow \qquad n^2+n-6=0
\end{eqnarray}
with the solutions
\beq
n_1=-3 \qquad {\rm{and}} \qquad n_2=2.
\eeq
The solution is a linear combination, giving $\Phi_2(r)=C_1 r^{-3}+C_2r^2$. 
Following \cite{Hinderer:2007mb} the integration constants $C_1$ and $C_2$ are 
\begin{equation} \label{one}
 C_1= \frac{15}{8M^3}\lambda \epsilon_{ij} \qquad {\rm{and}} \qquad C_2= \frac{M^2}{3} \epsilon_{ij}
\end{equation}
with some constant $\lambda$, whose physical interpretation will become important in the following. Now 
\begin{equation}\label{diff_eq_del_cazz_lin_comb}
\Phi_2(r)=\frac{8}{5}\left(\frac{M}{r} \right)^3 C_1+3\left(\frac{r}{M}\right)^2C_2.
\end{equation}
Inserting the integration constants from eq.~(\ref{one}) in eq.~(\ref{diff_eq_del_cazz_lin_comb}) gives
\begin{equation}
 \Phi_2(r)=\frac{3 \lambda}{2 r^3}\epsilon_{ij}+\frac{r^2}{2}\epsilon_{ij} \qquad {\rm{and}} \qquad  \Phi_2'(r)=-\frac{9 \lambda}{2 r^4}\epsilon_{ij}+r\epsilon_{ij}
\end{equation}
solving for $\epsilon_{ij}$ and arranging the terms yields
\begin{equation}
-9\lambda+2r^5=\frac{r\Phi_2'(r)}{\Phi_2(r)}(3\lambda+r^5).
\end{equation}
We substitute
\begin{equation}\label{wheyallstars}
y=\frac{r\Phi_2'(r)}{\Phi_2(r)}
\end{equation}
and find the tidal polarizability parameter $\lambda$ to be
\begin{equation}
\lambda=-\frac{(y-2)}{3(y+3)}R^5.
\end{equation}
Using eq.~(\ref{ruchnefischali}) we determine the so called Love number
\begin{equation}\label{kaa}
 k_2=\frac{1}{2}\frac{(2-y)}{(3+y)}.
\end{equation}
It is not surprising, that this equation can also be found for the Newtonian limit taken in the  Refs.~\cite{Love:1908tua,Hinderer:2007mb,Hinderer:2009ca,Postnikov:2010yn} where mainly general relativistic applications are discussed.\\
Because the density in every star depends on its radius and can of course not be assumed constant, our first result $k_2=3/5$ in 
eq.~(\ref{fleischgewordenerhass}) is replaced by eq.~(\ref{kaa}). The density dependence is hidden in the quantity $y$ in eq.~(\ref{wheyallstars}) which itself is a rational measure of the change of the gravitational potential quadrupole term divided by the quadrupole term itself. The quantity $y$ yields information about the deformability of the star, either due to rotation or due to the existence of an external perturbing tidal field. It is clear that $y$ is different for stars differently obtained, so that $y$ can be backtraced to the composition of the star, i.e. the equation of state.\\ 
Figure \ref{kzwoalsfvonypsilani} shows the Love number as function of the substituted parameter $y$. It can be seen that the root of eq.~(\ref{kaa}) is found at $y=2$. No change in $\Phi_2$ implies the star is not able to deform, $y \rightarrow 0$ and hence also $k_2 \rightarrow 0$. $y(r=0)=2$ is hence  the staring value to solve for a solution of $y$. That is because in the very center of the star at $r=0$ not deformability takes place.
\begin{figure}[H]
\begin{center}
\includegraphics[width=.6\linewidth,height=.4\linewidth]{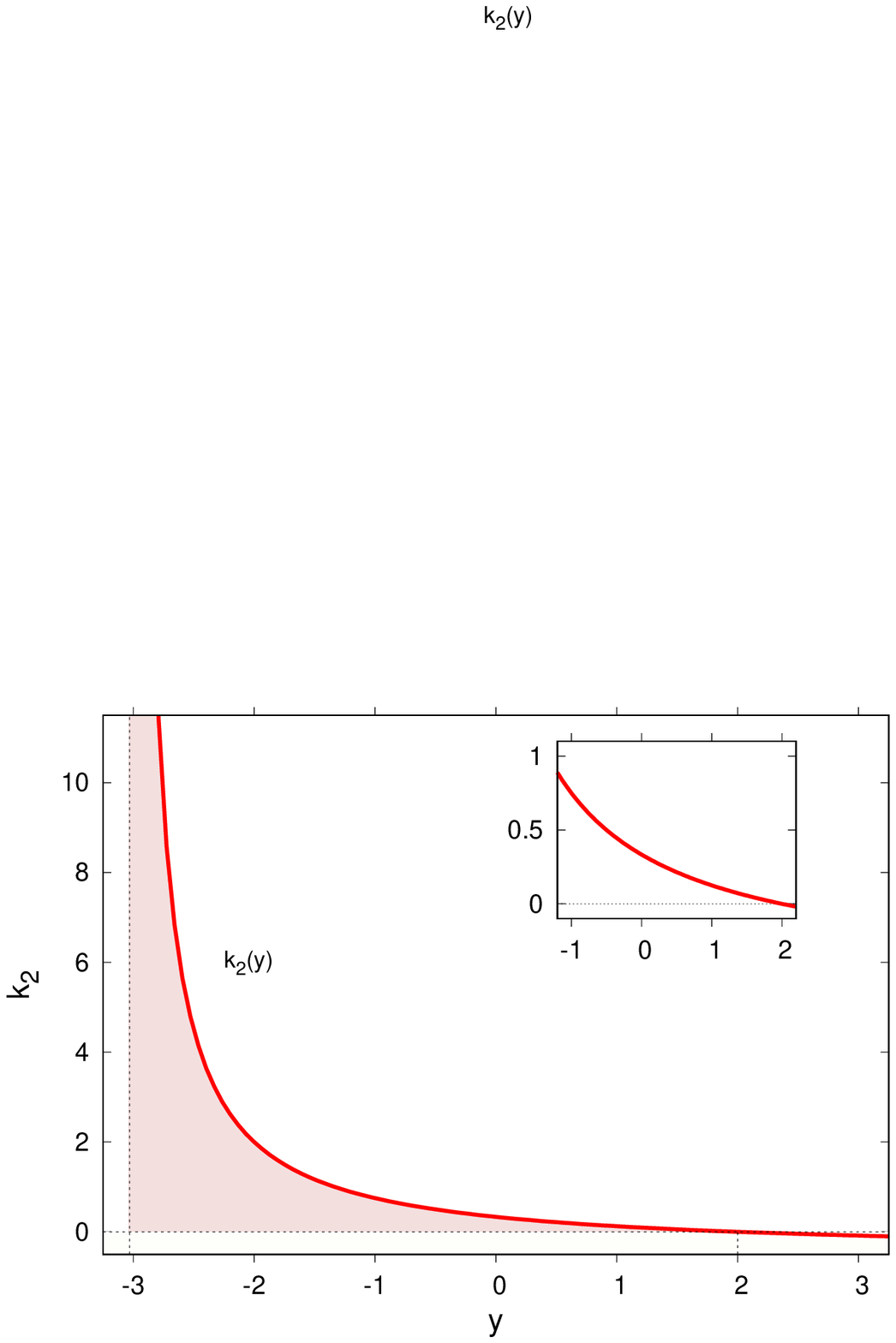}
\caption{\textit{$k_2$ as a function of $y$, see eq.~(\ref{kaa}). The solution of the differential equation eq.~(\ref{diff_eq_del_cazz}) allows only for values $-3\leq y \leq 2$, because $k_2 \geq 0$ for a physically reasonable deformation. A polytropic EoS with polytropic index $n=0$ allows for the smallest value of $y=-1$, which gives the upper bound on the Love number $k_2=0.75$, see section \ref{karmageddon}. For $n=0$ the star configurations are incompressible, explaining this feature of an upper bound.}}
\label{kzwoalsfvonypsilani}
\end{center}
\end{figure}
\subsection{The Love number for different Polytropes}\label{karmageddon}
For stars with nonzero density at the surface, such as selfbound stars  \cite{Ivanenko:1965dg,Itoh:1970uw,Bodmer:1971we,Haensel:1986qb,Alcock:1986hz,Weber:2004kj,Zacchi:2015lwa,Zacchi:2015oma} or stars resulting from a polytropic EoS with $n=0$, one has to substract an extra expression from eq.~(\ref{wheyallstars}) 
\begin{equation}\label{wheyallstarsmitextraspast}
y=\frac{r\Phi_2'(r)}{\Phi_2(r)}-\frac{4 \pi r^3 \Delta\epsilon}{m(r)},
\end{equation}
see also \cite{Hinderer:2007mb,Damour:2009vw,Hinderer:2009ca,Postnikov:2010yn}. 
The argument for selfbound stars is that because of the resulting discontinuity in the energy-mass density at the star's surface, $\epsilon$ jumps to zero. Because of this discontinuity, the value of $y(r)$ needs to be shifted according to eq.(\ref{wheyallstarsmitextraspast}).\\
Eq.~(\ref{hillary_clinton}) with its corresponding substitutions eq.~(\ref{substituete_1}) and eq.~(\ref{substituete_2}) for nonzero $U(r)$ and $V(r)$ gives the correct Newtonian approximation for the differential equation, 
\beq \label{billy_clinton}
\Phi_2''(r)+\frac{2}{r}\Phi_2'(r)+\left[\frac{4 \pi \rho}{dp/d\rho} - \frac{6}{r^2}\right]\Phi_2(r)=0
\eeq
which for instance is used in \cite{Hinderer:2009ca}. We turn now to discuss two cases of solutions of eq.~(\ref{billy_clinton}) for polytropes.
\begin{enumerate}
\item  For a polytropic EoS with $n=0$, which corresponds to an incompressible fluid $dp / d\rho \rightarrow \infty$ and the solution of  
       eq.~(\ref{billy_clinton}) for 
       $y(r)$ is simply a constant: The root of $k_2$ corresponds to the boundary condition $y(r=0)=2$ in the center of the star throughout the whole star $y(r)=2$. Because of the resulting discontinuity in the speed of sound $c_s^2=dp / d\rho=\infty$ at the surface, the extra term from eq.~(\ref{wheyallstarsmitextraspast}) has to be taken into account. The value of $y(r)$ needs to be shifted. Inserting the corresponding values gives 
       \beq \label{ichbineinenull}
       y_R=2-3=-1 \qquad {\rm{and}} \qquad k_2=3/4.
       \eeq
       The value $k_2=3/4$ corresponds to the upper limit, even when considering general relativistic effects.\\ 
The explanation for these quantities is that an incompressible fluid is the best fluid reacting to an external tidal field, or to rotation. 
In other words, the bulge at the equator due to rotation is largest for an incompressible fluid compared to compressible fluids. 
This fluid can not be compressed, therefore it is squished and flows in reaction to the external quadrupole field, like a balloon if you like. 
If on the contrary matter can be compressed, it forms a high density core and a low density mantle
around the core, more like a sponge if you like. The fluid can react to an external potential by increasing
the energy density in the core such that only a small quadrupole moment
is induced. By determining the Love number one learns something
about the compressibility of the fluid, and that translates into the softness or stiffness of the
EoS. The Love number $k_2$ gives most important information about the inner structure of stellar objects.\\
Interesting to note is that in case of polytropes the value of $k_2$ is independent on the value of the polytropic constant K (eq.~(\ref{Katznklo})), even if the maximum mases obtained depend on K. This feature can be explained via the scaling behaviour studied via dimensionless quantities: The corresponding radius scales with the mass. We will discuss this feature in the following section \ref{landsackschlack}.
\item  For the case of a polytrope with $n=1$ one finds that eq.~(\ref{billy_clinton}) is not that easy to solve compared to $n=0$. 
However, for $n=1$ the radius dependent density drops out in eq.~(\ref{billy_clinton}) so that an analytical solution is nonetheless possible.  With $p=K\rho^2$ eq.~(\ref{billy_clinton}) becomes 
\begin{eqnarray} 
\Phi_2''(r)+\frac{2}{r}\Phi_2'(r)+\left[\frac{2 \pi}{K} - \frac{6}{r^2}\right]\Phi_2(r)=0
\end{eqnarray}   
With the substitution $r=a \tilde r$ and the corresponding replacement of the function $\Phi_2(r) \rightarrow \Phi_2(\tilde r)$ and its derivatives 
\begin{eqnarray} \label{nochnittbessel}
\Phi_2''(\tilde r)+\frac{2}{\tilde r}\Phi_2'(\tilde r)+\left[\frac{2 \pi a^2}{K} - \frac{6}{\tilde{r}^2}\right]\Phi_2(\tilde r)=0
\end{eqnarray}  
We choose $a^2=K/2 \pi$ so that eq.~(\ref{nochnittbessel}) already resembles the solution of a Bessel function very closely. 
\begin{eqnarray} \label{nochnittbesseldenekste}
\underbrace{\Phi_2''(\tilde r)+\frac{2}{\tilde r}\Phi_2'(\tilde r)+\left[1 - \frac{6}{\tilde{r}^2}\right]\Phi_2(\tilde r)=0}_{\rm{our}\, \rm{finding}} \qquad \Longleftrightarrow  \qquad \underbrace{\Phi_2''(\tilde r)+\frac{1}{\tilde r}\Phi_2'(\tilde r)+\left[1 - \frac{p^2}{\tilde{r}^2}\right]\Phi_2(\tilde r)=0.}_{\rm{Bessel}}
\end{eqnarray}  
We are already familiar with substitutions, so that the factor two in front of $\Phi'(\tilde{r})$ can be treated with $\Phi=\tilde{\Phi}/\sqrt{r}$. After deriving $\tilde{\Phi}_2$ and after some non-enlightening algebra one can identify 
\begin{eqnarray} \label{dasaja}
\tilde{\Phi}_2''(\tilde r)+\frac{1}{\tilde r}\tilde{\Phi}_2'(\tilde r)+\left[1 - \frac{25}{4\tilde{r}^2}\right]\tilde{\Phi}_2(\tilde r)=0
\end{eqnarray}  
with the Bessel function $J_{5/2}$ for $p^2=25/4$. 
Resubstituing the corresponding derivatives of $\tilde{\Phi}_2$ and solving 
for $y$ in eq.~(\ref{wheyallstars}) 
one finds that
\beq \label{sporca_gudgiola}
y=\frac{-(p+\frac{1}{2})J_{5/2}+rJ_{3/2}}{J_{5/2}}=-3+\frac{r J_{3/2}}{J_{5/2}}.
\eeq
To evaluate the fractional relation of the Bessel functions in eq.~(\ref{sporca_gudgiola}) 
one may need the following relations for spherical Bessel functions
\begin{subequations}
 	\begin{equation}  J_{n+\frac{1}{2}}=\sqrt{\frac{2r}{\pi}}j_n  \end{equation}
 	\begin{equation}  j_n=r^n\left( \frac{1}{r}\frac{d}{dr}\right)^n\left( \frac{\sin(r)}{r}\right)  \end{equation}
 	\begin{equation}  j_{n+1}=\frac{n}{r}j_n-j'_1  \end{equation}
 	\begin{equation} J_{n+1}+J_{n-1}=\frac{2p}{r}J_n\end{equation}
\end{subequations}
It follows for 
\beq
J_{3/2}=\frac{r}{3}\left[J_{5/2}+\frac{\sin(r)}{r} \sqrt{\frac{2r}{\pi}} \right]
\eeq
which enters in eq.~(\ref{sporca_gudgiola}). With $J_{5/2}=\sqrt{\frac{2r}{\pi}} \cdot j_2$ and $j_2=\frac{1}{r}j_1-j'_1$ we find for $r=\pi$ that
\beq\label{mowgli}
y=-3+\frac{\pi^2}{3} \qquad {\rm{and}} \qquad k_2=(15-\pi^2)/2\pi^2.
\eeq
This is exactly the same result as taken for the Newtonian limit in Refs.~\cite{Hinderer:2007mb,Damour:2009vw,Hinderer:2009ca,Postnikov:2010yn}.
\end{enumerate}
For every other polytropic index n eq.~(\ref{billy_clinton}) has to be solved numerically because the density $\rho(r)$ gives a nontrivial contribution to the solution of eq.~(\ref{billy_clinton}).\\
The Love number encodes information about the star’s degree of central condensation \cite{Batygin:2009ps,Becker:2013fma}. Stars that are more centrally
condensed will have a smaller response to a tidal field, resulting in a smaller Love
number. These features translate into the EoS. 
\begin{figure}[H]
\begin{center}
\includegraphics[width=.6\linewidth,height=.4\linewidth]{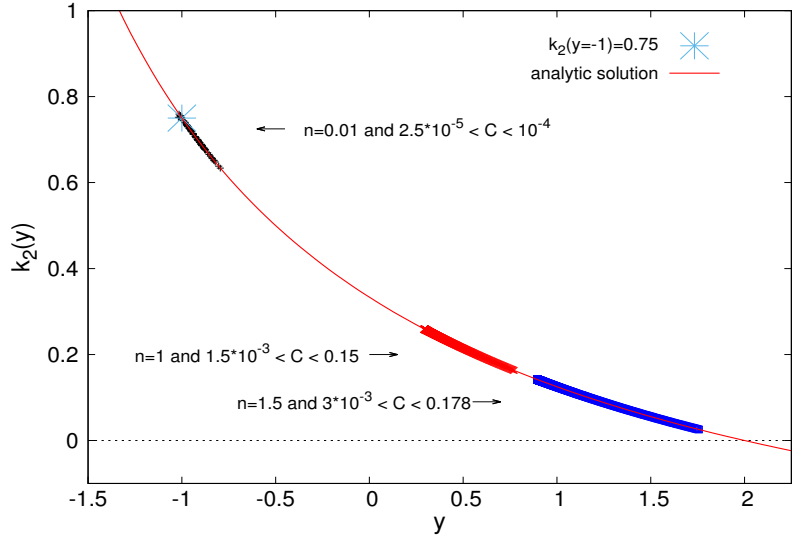}
\caption{\textit{$k_2$ as a function of $y$ for different polytropic indices n, according to eq.~(\ref{kaa}). It can be seen that for small $n=0.01$ and vanishing compactness the value of $k_2$ is determined only via $k_2(y=-1)=0.75$, i.e. is indeed the upper limit for an incompressible fluid. The larger n becomes the smaller is $k_2$, which implies that a stiffer EoS (smaller polytropic index n) yields stars which are less deformable due to the formation of a high density core and a rather less dense mantle around the core compared to stars obtained with a softer EoS.}}
\label{kzwoalsfvonypsilani}
\end{center}
\end{figure}
Figure~\ref{kzwoalsfvonypsilani} shows the analytic function of the Love number $k_2$ as a function of $y$ according to eq.~(\ref{kaa}). Numerical solutions for different polytropic indices are shown for $n=0.01$, $n=1.0$ and $n=1.5$. It can be seen that for relatively small compactness the solutions start at lower value of $y$ implying a larger value of $k_2$ compared to larger compactness. That means that less compact stars are more deformable due to a tidal field for a given value of the polytropic index n. With larger compactness the stars become less deformable, since the value of $k_2$ becomes smaller.\\ 
In other words: More condensed stars obtained with increasing polytropic index n are more compressible and have a smaller response to the perturbing tidal field, i.e. a rather small value of the Love number $k_2$.

\section{General Relativistic Physics}\label{gr_baby}
One of the reasons compact stars have to be described by general relativity (GR) is because of the enormous curvature effects on spacetime. Gravity crushes matter in compact stars to realms that lie far beyond a nonrelativistic treatment via classical physics. 
The detection of gravitational waves (GW) from a black hole-black hole merger in 2015 \cite{Abbott:2016blz,Abbott:2016nmj,TheLIGOScientific:2016pea} and from a binary neutron star merger in 2017 \cite{TheLIGOScientific:2017qsa,Abbott:2018wiz} clearly makes compact star physics not only interesting, but of actual interest.\\ 
Constraints on compact stars and hence on the underlying EoSs arise from 
the measurements of the masses of the pulsar PSR J1614-2230 in 2010 
and of PSR J0348+0432 in 2013 \cite{Demorest:2010bx,Antoniadis:2013pzd,Fonseca:2016tux,Cromartie:2019kug}. These stars with
$M \simeq 2~M_{\odot}$ for instance impose constraints on the equation of
state (EoS) for compact stars. Many EoSs describing compact objects 
could be ruled out since the corresponding mass radius relation does not reach this limit.
Still it is an open discussion if exotic matter is present in the core of such an object
\cite{Kaempfer:1981a,Kaempfer81,Kaempfer82,Kaempfer83a,Kaempfer83b,Kaempfer85,Glendenning:1998ag,Schertler:2000xq}. 
In several works it is also discussed if the whole object is  made of exotic matter such as a Kaon condensate or pure quark matter \cite{zdunikhaenselschaeffer:1983pti,Sagert:2008uq,SchaffnerBielich:2010am,Weissenborn:2011ut,Zacchi:2015lwa,Zacchi:2015oma}. 
A possible phase transition is cutting edge for twin star solutions \cite{Alford:2014dva,Alford:2015dpa,Benic:2014jia,Blaschke:2015uva,Zacchi:2016tjw,Christian:2017jni}. The inner composition of such stars is still poorly understood, not to say rather unknown.\\ 
The detection of the gravitational wave signal from the inspiral of two compact objects, which happened to take place
on August 17, 2017 \cite{TheLIGOScientific:2017qsa,Abbott:2018wiz} imposes another constraint on the zoo of EoSs. The GW signal is strongly related to the star's tidal deformability, which is linear proportional to the Love number of the star.

\subsection{Relativistic structure equations}\label{landsackschlack}
The equation for mass conservation eq.~(\ref{mass_appeal_madness}) also holds relativistically. Only, the mass density $\rho(r)$ is replaced with the energy-mass density $\epsilon(r)$ with $\rho=\epsilon/c^2$ where $c^2$ is the squared speed of light, so that
\beq \label{skyrim}
\frac{dm}{dr}=\frac{4 \pi r^2 \epsilon(r)}{c^2}
\eeq
The equations of Newton already tell us how mass acts as a source of the gravitational field
\beq
m\frac{dr^2}{dt^2}=-m\vec{\nabla}\Phi(\vec{r}) \qquad {\rm{and}} \qquad \vec{\nabla}\Phi(\vec{r})=4\pi G\rho(\vec{r}) 
\eeq
with $\Phi(\vec{r})$ as the gravitational potential, see also eq.~(\ref{source_of_wurst}). 
From special relativity it is known that energy and mass are equivalent. 
That means that energy is also a source of a gravitational field.
The equations have to become tensorical: The field $\Phi(\vec{r})$ is replaced by the metric $ g_{\mu \nu}$ and the corresponding counterpart on the other side of the equation is the energy momentum tensor $T_{\mu \nu}$.\\
The Einstein field equations $G_{\mu\nu}$ can be derived via the metric $g_{\mu\nu}$, which models the gravitational potential, and the Christoffel symbols, which model gravity as a pseudoforce. The Ricci tensor $R_{\mu\nu}$ models the curvature and is defined in terms of the  Christoffel symbols. The Einstein equations can be found in almost every textbook concerning general relativity and read
\begin{equation}\label{einigleini}
G_{\mu\nu}=R_{\mu\nu}-\frac{1}{2}g_{\mu\nu} R=\frac{8\pi G}{c^4}T_{\mu\nu},
\end{equation}
see for instance Ref.~\cite{Zeldovich_book,Glendenning_book,fliessbach_book,Stannard_book,hartle_book,wheeler_book}.
To ensure energy and momentum conservation, $T_{\mu\nu}$ has to be divergenceless. 
The Ricci tensor however is not divergenceless. Therefore the Einstein equations contain a
correction in form of a product of the metric and the Ricci scalar R. 
Matter, energy and spacetime influence each other, therefore the field equations are highly non-linear and exceedingly complicated to solve.
There are but a few solutions in closed form, of which the Schwarzschild solution the most famous one is. The Schwarzschild solution is a vacuum solution, i.e. $T_{\mu\nu}=0$, and it describes the metric outside a spherically symmetric and static star. For the interior structure of a spherically symmetric static star, the Tolman-Oppenheimer-Volkoff (TOV) equations, need to be solved. We will sketch the derivation in the following.\\ 
After a massive star ($M_{star}\geq 8 M_{\odot}$) has consumed its nuclear fuel, it eventually explodes in a Supernova Type II. The remnant is called a compact star.
The TOV equations describing such stars have first been derived in 1939 by Tolman, Oppenheimer and Volkoff \cite{Tolman39}.
For isotropic, static and spherically symmetric mass distributions and metric, the energy momentum tensor $T_{\mu\nu}$ adopts the hydrodynamical form, since the matter within a compact star can be treated as a relativistic fluid \footnote{Note that no electrical or magnetical fields are considered.}.
\beq
T_{\mu\nu}=\left(\rho+\frac{p}{c^2}\right)u_{\mu}u_{\nu}-pg_{\mu\nu}
\eeq
For the star to be in hydrodynamical equilibrium: $u^{\mu}=(u^0,\vec{0})$ and $g_{00}(u^0)^2=c^2$, from which follows that
$(u_0)^2=c^2g_{00}$. The energy momentum tensor then reads 
\begin{eqnarray} \label{ideal_fluid}
T_{\mu\nu} &=& \left(\rho+\frac{p}{c^2}\right)c^2g_{00}\delta_{\mu 0}\delta_{\nu 0}-pg_{\mu\nu}
=\left(\begin{array}{cccc}\rho_0c^2e^{\nu(r)}&0&0&0\\
0&pe^{\lambda(r)}&0&0\\
0&0&pr^2&0\\
0&0&0&pr^2\sin^2{\theta}
\end{array}\right)
\end{eqnarray}
with $e^{\nu(r)}$ and $e^{\lambda(r)}$ as the metric functions from GR, see also eq.~(\ref{laendinnhuhn}).
The condition for hydrodynamic equilibrium requires that the covariant derivative 
of the enery momentum tensor vanishes, which eventually leads to 
\beq\label{depp_der}
\frac{dp}{dr}=-\frac{1}{2}\nu'(r)\left(\rho(r)c^2+p(r)\right)
\eeq
$\nu'(r)$ has to be determined by the Einstein eqs.~(\ref{einigleini}) :
\begin{eqnarray}
G_{tt}&=&e^{\nu - \lambda}\left(-\frac{\lambda'}{r}+\frac{1}{r^2}\right)-\frac{e^{\nu}}{r^2}=-\frac{8\pi G}{c^4}\rho e^{\nu}\\
G_{rr}&=&-\frac{\nu'}{r}-\frac{1}{r^2}+\frac{e^{\lambda}}{r^2}=-\frac{8\pi G}{c^4}p e^{\lambda}\\
G_{\Phi\Phi}&=&-\frac{1}{2}r^2e^{-\lambda}\left( \nu''+\frac{\nu' \lambda}{2}- \frac{\nu' \lambda'}{2} +\frac{\nu' - \lambda'}{r}\right)=-\frac{8\pi G}{c^4}pr^2\\
G_{\theta\theta}&=&\sin^2(\theta)G_{\Phi\Phi}=-\frac{8\pi G}{c^4}pr^2\sin^2(\theta) 
\end{eqnarray}
Solving for $\nu'$ and substituting into eq.~(\ref{depp_der}) yields 
\begin{eqnarray} \label{tov_1}
\frac{dp}{dr}&=&-\frac{(\rho c^2+p)\left[G m(r)+\frac{4\pi G}{c^4}p(r)r^3 \right]}{r^2-2m(r)r}\\ \nonumber
&=&-\frac{G \epsilon(r)m(r)}{(cr)^2}\left(1+\frac{p(r)}{\epsilon(r)}\right)\left(1+\frac{4\pi r^3p(r)}{m(r)c^2}\right)\left(1-\frac{2 m(r)}{c^2r}\right)^{-1}
\end{eqnarray}
It can be seen that for $m(r)\ll r$ and $p \ll \rho c^2$
the equation reproduces the Newtonian limit, eq.~(\ref{hydro_schraubenschluessel}). Mass conservation, eq.~(\ref{skyrim}), is valid without any corrections.
Eqs.~(\ref{skyrim}) and (\ref{tov_1}) are called the TOV equations.
The key differences between stellar Newtonian gravity and general relativity can be attributed to the differences from eq.~(\ref{hydro_schraubenschluessel}) to eq.~(\ref{tov_1}) 
\begin{itemize}
\item [(i)] Gravity does not only act upon the the density $\rho$ but also on the pressure 
      p, i.e. $\rho c^ 2 \rightarrow \rho c^2+p$ in the first bracket of eq.(\ref{tov_1}).
\item [(ii)] The second bracket in eq.(\ref{tov_1}) tells us that not only the mass, but 
      every single entry from $T_{\mu\nu}$ contributes to the gravitational field,
      i.e. energy is also a source of gravity, $m(r)\rightarrow m(r) + 4\pi r^3 p(r)$
\item [(iii)] The third one reflects the difference between Newtonian gravity and the 
      gravitational ``force'' due to general relativity. This accounts 
      for the additional factor in the denominator. 
\end{itemize}
All these terms come along with a factor $\propto c$ relative to the leading Newtonian terms.
The TOV equations are usually solved numerically with an appropriate EoS, typically given by a relation between pressure $p$ and energydensity $\epsilon(r)$. The boundary conditions are $m(r=0)=0$ and $p(r=0)=p_c$. For each EoS exists a 
solution which is parametrized by $p_c$, the central pressure of the star. 
The entire collection of masses and corresponding radii is called the mass-radius relation of compact stars.\\
Interesting to discuss in our analysis are scaling solutions of the TOV equations. A particular kind of such a scaling solution was already adopted for the study of the Lane Emden equations discussed in section~\ref{Lahme_Ente}. However, to the purpose of scaling one sets $c=1$ in eq.~(\ref{tov_1}) and introduces dimensionless quantities for the EoS $p(\epsilon)$.
\begin{eqnarray}
 P=\epsilon_0 P' \qquad {\rm{and}}  \qquad \epsilon=\epsilon_0 \epsilon'
\end{eqnarray}
where the primed quantities denote the dimensionless quantities and $\epsilon_0$ is a typical energy scale. Introducing further 
\begin{eqnarray}
 r=\mathcal{R} r' \qquad {\rm{and}}  \qquad m(r)=\mathcal{M} m'(r)
\end{eqnarray}
with a typical mass $\mathcal{M}$ and a typical radius $\mathcal{R}$ and plugging everything in eq.~(\ref{tov_1}) 
one arrives at 
\begin{eqnarray}
 \mathcal{R}=\frac{1}{\sqrt{G \epsilon_0}} \qquad {\rm{and}} \qquad  \mathcal{M}=\frac{1}{G}\frac{1}{\sqrt{G \epsilon_0}}
\end{eqnarray}
The statement here is that mass and radius scale with a constant factor $\epsilon_0$ 
to some power for a given EoS of the form $P' = P'(\epsilon')$. 
From this analysis we can immediately write down how the maximum mass $\mathcal{M}_{max}$ and the corresponding radius $\mathcal{R}_{crit}$ depend on energy density scale $\epsilon_0$.
\begin{eqnarray}\label{scaleyoubitch}
 \mathcal{R}_{crit}=\frac{\mathcal{R}'_{crit}}{\sqrt{G \epsilon_0}} \qquad {\rm{and}} \qquad  \mathcal{M}_{max}=\frac{\mathcal{M}'_{max}}{\sqrt{G^ 3 \epsilon_0}}
\end{eqnarray}
This feature implies that a relation such as the compactness $C \propto M/R$ is scale invariant.

\subsection{General relativistic treatment of tidal effects}
Apart from the two solar mass limit \cite{Demorest:2010bx,Antoniadis:2013pzd,Fonseca:2016tux,Cromartie:2019kug}, another constraint on the zoo of EoSs comes from the 
detection of the gravitational wave signal from the inspiral of two compact objects \cite{Radice:2017lry,Margalit:2017dij,Rezzolla:2017aly,Abbott:2018exr,Most:2018hfd,Annala:2017llu,De:2018uhw,Kumar:2017wqp,Fattoyev:2017jql,Malik:2018zcf}. This event happened to take place
on August 17, 2017 \cite{TheLIGOScientific:2017qsa,Abbott:2018wiz}. As the stars orbit each other they loose energy due to the emission of gravitational waves and as the orbit of this system shrinks, the inspiral accelerates. In the final minutes before the stars collide, this signal is strong enough to be detectable for ground based detectors. 
The component masses of the binary neutron star merger (BNSM) are in the range $1.16 \leq M_{\odot} \leq 1.6$, with the total mass of the system $2.73^{+0.04}_{-0.01}M_{\odot}$.\\ 
It is possible that a phase perturbation of the GW signal gives 
information about the compact star structure \cite{Flanagan:2007ix,Hinderer:2007mb,Hinderer:2009ca}. 
The influence of the internal structure on the
GW signal of the inspiral is characterized by a
single parameter $\lambda$, which is the ratio of the induced quadrupole $Q_{ij}$ to the perturbing
tidal field $\epsilon_{ij}$ from the companion. $\lambda$ is called the tidal polarizability
\beq
Q_{ij}=-\lambda \epsilon_{ij}=-\frac{2k_2 R^5}{3G} \epsilon_{ij}
\eeq
where $k_2$ is the star's dimensionless tidal Love number \cite{Love:1908tua} and $R$ is the radius of the star, see also eq.~(\ref{ruchnefisch}). 
We have already discussed that more condensed stars have a smaller response to the perturbing tidal field which will result in a smaller $k_2$. According to \cite{Hinderer:2009ca} neutron stars differ
in the value of $k_2$ from the Newtonian values up to $24\%$ using polytropic models, which are poor approximations for relativistic compact stars but nonetheless giving direction.
The tidal deformability parameter $\Lambda$ depends on the compactness $C$ of the compact star and on the Love number $k_2$ \cite{Love:1908tua,Hinderer:2007mb,Postnikov:2010yn} via 
\begin{equation}\label{loveandloveismierda}
 \Lambda=\frac{2 k_2}{3 C^5}.
\end{equation}
and is the most common value when discussing constraints on the EoSs. The compactness will become important when discussing tidal effects in GR. 
The GW170817 measurement on the tidal deformability using a 90\% probability density interval 
deduces $\Lambda=300^{+420}_{-230}$ for stars $1.16 \leq M/M_{\odot} \leq 1.60$ \cite{Abbott:2018wiz}. Inferred from that measurement, the radius of a 1.4$M_{\odot}$ compact star cannot be larger than $R \geq 13.6$~km \cite{Rezzolla:2017aly,Abbott:2018exr,Most:2018hfd,Annala:2017llu,De:2018uhw,Kumar:2017wqp,Fattoyev:2017jql,Malik:2018zcf,Hornick:2018kfi,Zacchi:2019ayh}.\\
\subsection{Quadrupole moments, Love number and tidal deformations}\label{mannomausinger}
In this section we briefly sketch how to derive relativistic tidal interactions. A complete treatment of the problem is beyond the scope of this article but can be found in \cite{Flanagan:2007ix,Hinderer:2007mb,Hinderer:2009ca,Postnikov:2010yn} and Refs. therein.
Starting point is the derivation 
of the general relativistic differential equation for the linear perturbations within the metric 
\beq
\tilde{g}_{\mu\nu}=g_{\mu\nu}+h_{\mu\nu}
\eeq
Following  \cite{1967ApJ...149..591T}, the angular dependence of the linear perturbation is described via spherical harmonics 
\beq
h_{\mu\nu}={\rm{diag}}[-e^{\nu(r)}H_0(r),e^{\lambda}(r)H_2(r), r^2K(r),r^2\sin(\theta)K(r)]Y_{2m}(\theta,\phi)
\eeq
It turns out that the function H(r) corresponds to the classically derived $\Phi_2$ as the solution of the differential equation, which has been derived classically in detail in section \ref{fressoderwatt}.
\begin{equation}\label{dasganzeding}
  H''(r)+H'(r)\left[\frac{2}{r}+e^{\lambda(r)}\left(\frac{2m(r)}{r^2}+4\pi r
   \left[p(r)-\epsilon(r)\right]\right)\right]+H(r)Q(r)=0
\end{equation}
where 
\begin{equation}
Q(r)=4\pi e^{\lambda(r)}\left(5\epsilon(r)+9p(r)+\frac{\epsilon(r)+p(r)}{c_s^2(r)}\right)-6\frac{e^{\lambda(r)}}{r^2}-(\nu'(r))^2
\end{equation}
with $c_s(r)^2=d p/ d \epsilon$ as the speed of sound squared and the metric functions from general relativity
\begin{equation} \label{laendinnhuhn}
e^{\lambda(r)}=\left(1-\frac{2m(r)}{r} \right)^{-1} \qquad {\rm{and}} \qquad \nu'(r)=2e^{\lambda(r)}\frac{m(r)+4\pi r^3 p(r)}{r^2}
\end{equation}
The solution of the differential equation~(\ref{dasganzeding}) describes the tidal perturbations of a compact star, derived within general relativity.
The Love number eventually is
\begin{eqnarray} \label{loveusorhateus}
k_2&=&\frac{8 C^5}{5} (1-2 C)^2 [2+2C(y_R-1)-y_R] \times \nonumber \\
&& \{ 2C[6-3y_R+3C(5y_R-8)]+\nonumber \\
&&4C^3[13-11y_R+C(3y_R-2)+ 2C^2(1+y_R)] + \nonumber \\
&&3 (1-2C)^2[2-y_R+2C(y_R-1)] \ln(1-2C) \}^{-1}  . \ \ \ \ \ \
\end{eqnarray}
Compared to the classical case, eq.~(\ref{kaa}), in GR $k_2$ 
depends not only on $y_R\equiv y(R)$ but also on the compactness $C=M/R$.  
This feature results from the deformation of spacetime due to GR and is directly related to the term 
\beq \label{nindalf_fennfeld}
e^{\lambda(r)}\left(\frac{2m(r)}{r^2}+4\pi r \left[p(r)-\epsilon(r)\right]\right)
\eeq
in eq.~(\ref{dasganzeding}). This can be seen when comparing eq.~(\ref{dasganzeding}) from GR with the classical terms in eq.~(\ref{hillary_clinton}) or eq.~(\ref{wheyallstarsmitextraspast}) for the potential. 
The solution of the differential equation for $H(r)$ can be simplyfied by casting eq.~(\ref{dasganzeding}) as a first order differential equation for
\begin{equation}\label{bueroklammer}
y_R\equiv y(R)=\frac{rH'(r)}{H(r)},
\end{equation}
which is straight forward but tedious to check. The similarity to the classical treatment in section \ref{bobgeldorf}, where the formalism has been derived classically, is apparently noteable, see eq.~(\ref{wheyallstars}). Finally  
\begin{eqnarray}\nonumber
 ry'(r)&+&y(r)^2+r^2 Q(r) \\ \label{dgl_del_czz}
 &+&y(r)e^{\lambda(r)}\left[1+4\pi r^2(p(r)-\epsilon(r) )\right] = 0
 ,
 \label{y}
\end{eqnarray}
see also \cite{1967ApJ...149..591T,Hinderer:2007mb,Damour:2009vw}.\\ 
This substitution circumvents the detour of solving for $H(r)$, when casting eq.~(\ref{dasganzeding}) as a first order differential equation using the same substitution for $H(r)$ as for $\Phi_2(r)$, see eq.(\ref{wheyallstars}). The difference is that one solves for $y$ directly and the value of $H(r)$ is irrelevant.
The boundary condition of eq.~(\ref{y}) is again $y(0)$=2, which implies no deformation at all in the center of the star, see also fig.~\ref{kzwoalsfvonypsilani}.    \\ 
If there is a discontinuity at the surface of the star at $r=R$, eq.~(\ref{bueroklammer}) has to be corrected via 
\begin{equation}\label{gr_wheyallstarsmitextraspast}
y=\frac{rH'(r)}{H(r)}-\frac{4 \pi R^3 \epsilon(r)}{M(r)}
\end{equation}
which we have already discussed within the classical treatement in Sec.~\ref{karmageddon}, see eq.~(\ref{wheyallstarsmitextraspast}).
The dimensionless tidal deformability $\Lambda$ in the general relativistic treatment is given by eq.~(\ref{loveandloveismierda}), which is 
usually solved simultaneously with
the TOV equations \cite{Hinderer:2007mb,Hinderer:2009ca,Postnikov:2010yn}. \\
Interesting to note is that the value of the compactness $C=0.5$ corresponds to a 
black hole implying $k_2=0$, i.e. such an object cannot be deformed by tidal disruptions.\\
\section{Results}
\begin{figure}[H]
\begin{center}
\includegraphics[width=.6\linewidth,height=.4\linewidth]{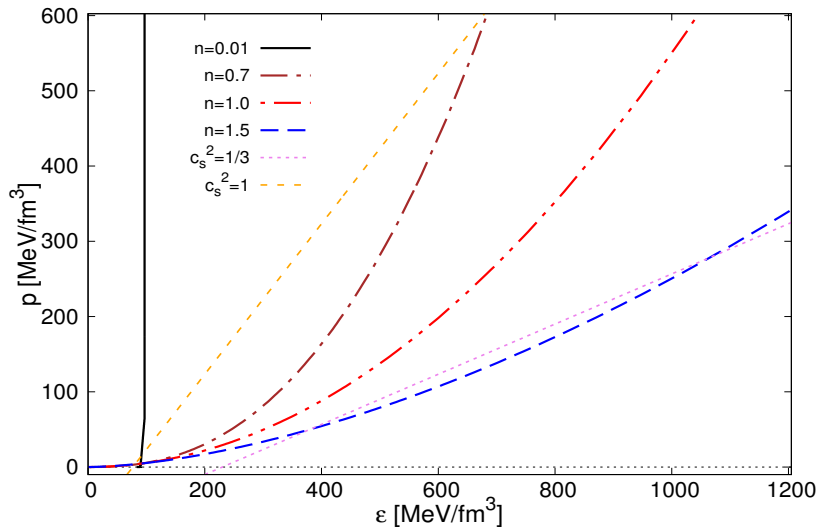}
\caption{\textit{The equations of state for different polytropic indices $n=0.01$, $n=0.7$, $n=1$ and $n=1.5$. The constant K has been chosen to be $K=\sqrt{24}~\rm{MeV/fm^3}$, which is around three times nuclear saturation density $n_0=0.16$~$\rm{fm^{-3}}$. It can be seen that for larger values of n, the EoS becomes softer, that is a smaller value of the pressure at a given energy density. The linear EoSs for $c_s^2=1/3$ and $c_s^2=1$ correspond to an ultrarelativistic gas \cite{Chodos74}, i.e. eq.~(\ref{selfdestruction}). Note, that for $p=0$ the energy density is not necessarily zero: The Bag constant $B^{1/4}$ has been chosen to be 145~$\rm{MeV/fm^3}$.}}
\label{hastduwurstinderkaraffe}
\end{center}
\end{figure}

\begin{figure}[H]
\begin{center}
\includegraphics[width=.6\linewidth,height=.4\linewidth]{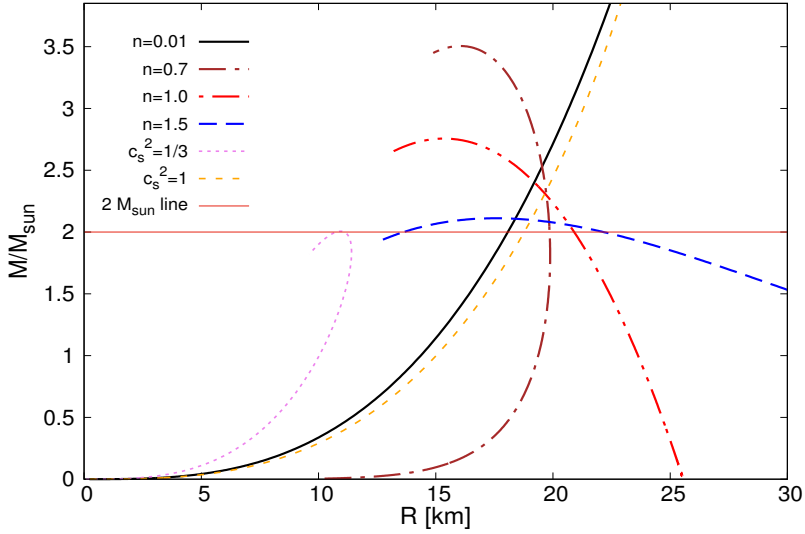}
\caption{\textit{The mass radius relations for the different polytropic EoSs shown in fig.~\ref{hastduwurstinderkaraffe}. For larger polytropic index n the maximum mass becomes smaller. The cases for $n=0.01$ and for $c_s^2=1$ are relatively close to each other, only that for a given mass the $c_s^2=1$ case yields larger values of the radius.  The radii for any given mass are the smallest for $c_s^2=1/3$, see also fig.~\ref{jonesy}. The $2 M_{\odot}$ limit is in any case fulfilled.}}
\label{ischblasdirdenmarschduarsch}
\end{center}
\end{figure}

Figure \ref{hastduwurstinderkaraffe} shows the equations of state (EoSs) for different polytropic indices $n=0.01$, $n=0.7$, $n=1$ and $n=1.5$. The constant $K$ for the polytropic EoSs have been chosen to be $K=\sqrt{24}~\rm{MeV/fm^3}$, which is around three times nuclear saturation density $n_0=0.16$~$\rm{fm^{-3}}$ according to eq.~(\ref{Katznklo}). This value has been chosen such that the softest EoS for $n=1.5$ still yields $2M_{\odot}$ due to the measurements of $2 M_{\odot}$ pulsars \cite{Demorest:2010bx,Antoniadis:2013pzd,Fonseca:2016tux,Cromartie:2019kug}.  
Figure \ref{hastduwurstinderkaraffe} also shows two EoSs for an ultrarelativistic gas according to eq.~(\ref{selfdestruction}), one for $c_s^2=1/3$ and the other one as limiting case due to causality, $c_s^2=1$. These two EoSs are particular, because of the non-vanishing value of the energy density at zero pressure due to the vacuum pressure constant B. We choose the common value of $B^{1/4}=145~\rm{MeV/fm^3}$ \cite{Schertler:2000xq} but the value of this constant depends on the model one studies, see for instance  \cite{Chodos74,Baym:1976yu,Aerts78,Schertler:1998cs,Klahn:2015mfa,Zacchi:2015lwa,Zacchi:2016tjw} and Refs. therein. The resulting stars of such  EoSs shown in the mass radius relation in figure~\ref{ischblasdirdenmarschduarsch} are so called selfbound stars. 
They do not need gravity to be stable \cite{Glendenning:1984jr}.
The smaller the polytropic index n, the stiffer is the corresponding EoS, that is a larger value of the pressure at a given value of the energy density. $n=0.01$ corresponds to the limiting case of eq.~(\ref{polytropic_eos}) for $p=K\epsilon^{\Gamma=\infty}$ for a polytropic index of $\lim n\rightarrow 0$. We choose $n=0.01$ because with a value of exacty zero one runs into numerical trouble.\\

\begin{figure}[H]
\begin{center}
\includegraphics[width=.6\linewidth,height=.4\linewidth]{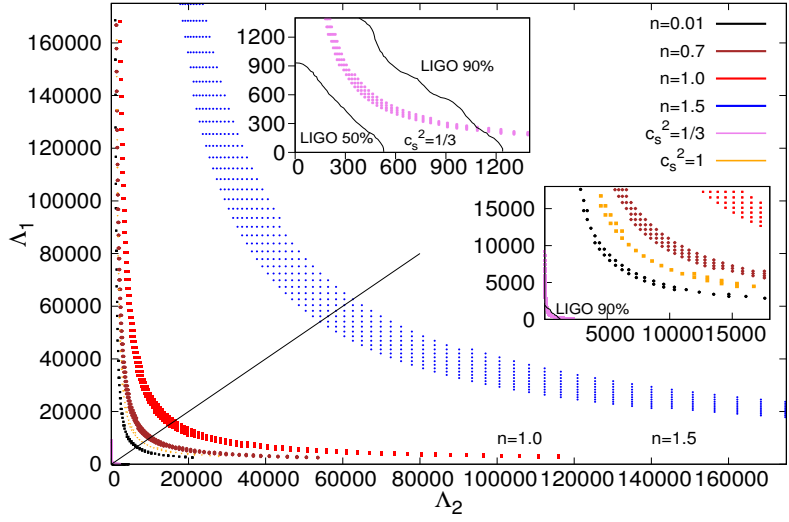}
\caption{\textit{$\Lambda_1$ vs. $\Lambda_2$ plot for different polytropic indices n and the two solutions of the constant speed of sound EoSs. The inlaid figures accentuate  better resoultions due to the large spreading of the results. Every result apart from the $c_s^2=1/3$ lies well outside of the 90\% credibility region by LIGO \cite{TheLIGOScientific:2017qsa}.}}
\label{leolausemaus}
\end{center}
\end{figure}

Figure \ref{ischblasdirdenmarschduarsch} shows the solutions of the TOV equations, eq.~(\ref{skyrim}) and eq.~(\ref{tov_1}). The mass radius relations are obtained for different EoSs seen in the previous figure \ref{hastduwurstinderkaraffe}. 
For larger polytropic index n the maximum mass becomes smaller. For $n=0.7$ $\rightarrow$ $M_{max} \simeq 3.5M_{\odot}$ at $R \simeq 16$~km, $n=1.0$ $\rightarrow$ $M_{max} \simeq 2.75M_{\odot}$ at $R \simeq 15.4$~km and $n=1.5$ $\rightarrow$ $M_{max} \simeq 2.1M_{\odot}$ at $R \simeq 17.5$~km.  
The mass radius relation for $n=0.01$ and for $c_s^2=1$ are relatively close to each other, only that for a given mass the $c_s^2=1$ EoS yields slightly larger values of the radius. This will become important when discussing the tidal deformability of two colliding stars in fig.~\ref{leolausemaus}. The masses (and radii)  obtained with these two EoSs are far to large to be physically reasonable. Their feature as a limiting case however makes them interesting nonetheless.

\begin{figure}[H]
\begin{center}
\includegraphics[width=.6\linewidth,height=.4\linewidth]{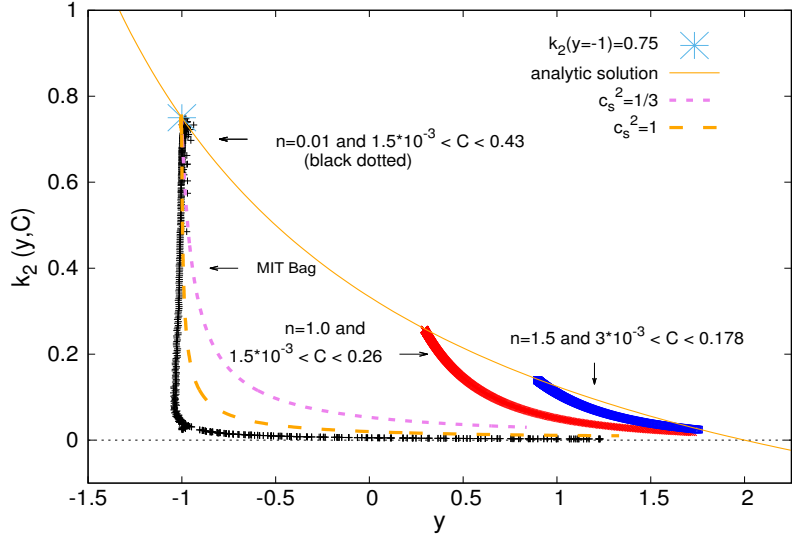}
\caption{\textit{$k_2$ depending on the compactness $C$ as a function of $y$ for different polytropic indices n according to eq.~(\ref{loveusorhateus}). It can be seen that for small $n=0.01$ and vanishing compactness the value of $k_2$ is determined only via $y=-1$, i.e. is indeed the upper limit for an incompressible fluid. The larger n becomes the smaller is $k_2$, which implies that a softer EoS yields stars which are more deformable compared to stars obtained with a stiffer EoS. This feature arises due to the formation of a high density core and a rather less dense mantle around the core. The ultrarelativistic cases for $c_s^2=1$ and $c_s^2=1/3$ are found closely to the incompressible polytrope implying that selfbound stars are rather hard to compress. The analytic solution corresponds to the classically derived quantity $k_2$, see eq.~(\ref{kaa}).}}
\label{kzwoalsfvonypsilaniundaltmeyer}
\end{center}
\end{figure}
The best candidate when comparing with observable quantities is the mass radius relation for an ultrarelativistic gas for a speed of sound $c_s^2=1/3$. The radii for any given mass are the smallest in our analysis. We will see that this feature is for favoured a tidal deformability parameter compliant with the value determined via the detection of GW. The $2 M_{\odot}$ are also reached at $R \simeq 11$~km.\\

Figure \ref{kzwoalsfvonypsilaniundaltmeyer} shows the Love number $k_2$ as a function of the compactness. Since in GR $k_2$ depends not only on the perturbation $y$, but also on the compactness, the results are different to those obtained classically in fig.~\ref{kzwoalsfvonypsilani}.\\ 
It can be seen that the results for a Polytrope with $n \rightarrow 0$ and very small compactness $C=1.5 \cdot 10^{-3}$ start at $k_2(y=-1) \simeq 3/4$, as is the case for the classical analysis of $k_2$. 
The solutions for $n=0.01$ drop nearly vertical from the analytic solution $k_2(y,C)=k_2(-1,0)=0.75$ according to eq.~(\ref{kaa}) for larger compactness of the stars. That is to say that the perturbation $y$ stays approximately constant at $y=-1$, see also Tab.~\ref{terence_hill}.\\ 
The physical interpretation is that for constant perturbation $y \simeq -1$, the deformability due to the influence of the Love number is mainly determined by the compactness $C$. At a certain compactness $k_2 \simeq 0$ and the compact object is (approximately) not deformable by tidal disruptions. This can also be understood with the argument of incompressibility discussed in sec.~\ref{karmageddon}. Explained vividly: A balloon which is only a little inflated can be deformed easily, but it is still incompressible. If you pump it up with air, you increase the pressure. This corresponds to a larger compactness and the balloon is hence harder to deform for the same squeezing. This feature is modelled via the quantity $y$.\\
However, the EoSs with a constant speed of sound show an nearly equal development, only $y$ increases slightly before $k_2$ reaches zero. $c_s^2=1$ is closer to the $n=0.01$ case as $c_s^2=1/3$. The polytropes for $n=1.0$ and for $n=1.5$ start at larger values of $y$, i.e. smaller values of $k_2$. 
Such fluids can react better to an external potential, for instance by increasing the energy density in the core. Every shell of matter passes the influence of the external potential to the next shell towards the center of the star. As a result only a small quadrupole moment is induced which yields a rather small Love number $k_2$. The case for $n=1$ has also been discussed classically in sec.~\ref{karmageddon}. The same statement as for classical mechanics is valid for GR: The larger the polytropic index n, the softer is the EoS and the smaller is $k_2$. This  implies that a stiffer EoS (when n becomes smaller) yields stars which are less deformable, because the matter becomes more and more incompressible. For increasing polytropic index n, a high density core and a rather less dense mantle around the core forms.
For very small compactness all solutions start at the classical value (that is the analytical solution in fig.~\ref{kzwoalsfvonypsilaniundaltmeyer}) because in classical mechanics a quantity such as the compactness does not come to play in the formalism. Solutions for $C \rightarrow 0$ can be calculated with the Newtonian limit of eq.~(\ref{loveusorhateus}), i.e. eq.~(\ref{kaa}).

\begin{table}[H]
\begin{center}
\begin{tabular}{|c|c|c|c|} 
\hline \hline
Polytropic index n  & $y=rH'/H$ &  $C \propto M/R$ & $k_2(y,C)$    \\ \hline
0.01 & $\simeq$ -1 & $1.5 \cdot 10^{-3}$  & 0.75 \\ \hline
0.01 & $\simeq$ -1 & 0.1  & 0.4 \\ \hline
0.01 & $\simeq$ -1 & 0.26  & 0.1 \\ \hline
0.01 & $\simeq$ -1 & 0.4  & 0.01 \\ \hline
\end{tabular}  
\caption{\textit{The numerical values of the Love number $k_2(y,C)$ for a polytrope with $n \rightarrow 0$. The more compact a star is, the harder it becomes to deform, which is expressed in a smaller value of $k_2$.}}
\label{terence_hill}
\end{center}
\end{table}
Generally one can say that a very compact \textit{compact star} for any EoS is harder to deform then a less compact one.  
This statement is intuitive (exceptions however are low-mass compact stars).  
For all these EoSs it is interesting to note that $k_2(y,C)$ is independent on the values of the constants in the EoSs. This holds for either polytropic or ultrarelativistic EoSs. This arises from the scaling behaviour which we have discussed at the end of sec.~\ref{landsackschlack}, see eq.~(\ref{scaleyoubitch}). This is also found numerically in  \cite{Postnikov:2010yn}.\\

\begin{figure}[H]
\begin{center}
\includegraphics[width=.6\linewidth,height=.4\linewidth]{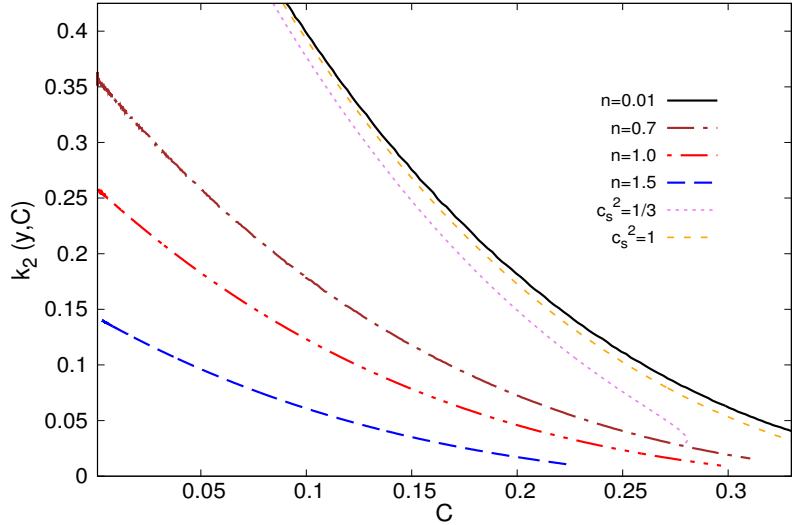}
\caption{\textit{$k_2$ depending on $y$ and on the compactness $C$ for different polytropic indices n according to eq.(\ref{loveusorhateus}). Generally, $k_2$ decreases with increasing polytropic index n and increasing compactness $C$. The ultrarelativistic EoSs with constant speed of sound are located closely to the polytropic case for $n=0.01$. The same figure can be found in Refs.~\cite{Hinderer:2009ca,Postnikov:2010yn} for instance.}}
\label{sostellstdudiraankidmnappingvor}
\end{center}
\end{figure}

Figure~\ref{sostellstdudiraankidmnappingvor} depicts $k_2$ as a function of $y$ and on the compactness $C$ for different polytropic indices n according to eq.~(\ref{loveusorhateus}).  Generally, $k_2$ decreases with increasing polytropic index n and increasing compactness $C$. For vanishing compactness $C \rightarrow 0$, i.e. within a Newtonian approach,  the values of $y$ and $k_2$ can be determined analytically for $n=0$, eqs.~(\ref{ichbineinenull}), and $n=1$, eqs.~(\ref{mowgli}), see section \ref{karmageddon}. 
The ultrarelativistic EoSs with constant speed of sound are located closely to the polytropic case for $n=0.01$. The case for $c_s^2=1$ is closer to $n=0.01$ then $c_s^2=1/3$. 
This again underlies the statements made before, see also fig.~\ref{kzwoalsfvonypsilaniundaltmeyer}.
The same figure \ref{sostellstdudiraankidmnappingvor} can also be found in Refs.~\cite{Hinderer:2009ca,Postnikov:2010yn}.\\

When two compact stars orbit each other they lose energy due to the emission of gravitational waves. The inspiral accelerates and the orbit shrinks. One star starts to deform due to the tidal field generated by the other star, both stars deform actually due to the other stars influence.  
The component masses of the binary neutron star merger (BNSM) detected in the GW170817 event are in the range $1.16 \leq M_{\odot} \leq 1.6$, with the total mass of the system $2.73^{+0.04}_{-0.01}M_{\odot}$ \cite{TheLIGOScientific:2017qsa,Abbott:2018exr}.
For stars in that mass range which are generated with the same EoS one can analyze the values of the tidal deformability in a $\Lambda$-$\Lambda$ plot and compare these results with the measurement by LIGO \cite{TheLIGOScientific:2017qsa,Abbott:2018exr}. The $\Lambda$ parameters characterize the size of the tidally induced mass deformations of each star and are proportional to $\Lambda \propto k_2 (R/M)^5$, see eq.~(\ref{loveandloveismierda}) and also 
Refs.~\cite{TheLIGOScientific:2017qsa,Abbott:2018exr} for more details.\\

\begin{figure}[H]
\begin{center}
\includegraphics[width=.6\linewidth,height=.4\linewidth]{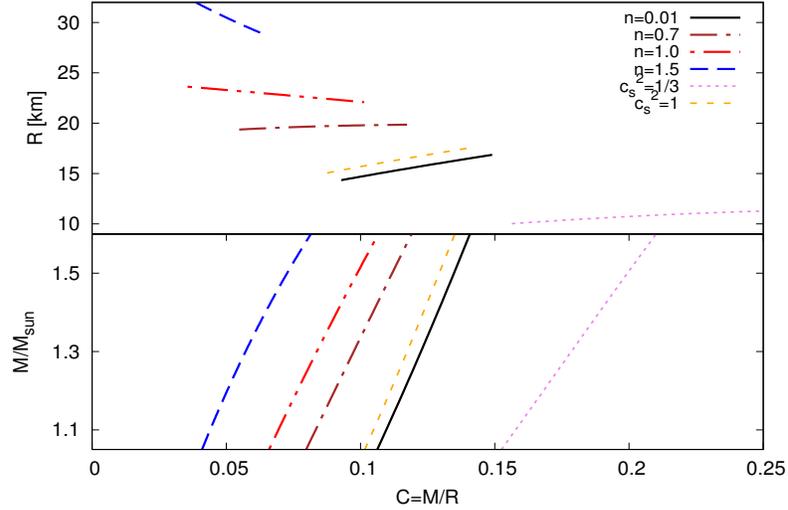}
\caption{\textit{The radius and the mass as a function of the compactness for the different polytropes and the constant speed of sound EoSs discussed. 
The radii are those of stars in a mass range $1.1 \leq M/M_{\odot} \leq 1.60$ range . These can collide to form a total mass of the proto neutron star $2.73^{+0.04}_{-0.01}M_{\odot}$ according to \cite{TheLIGOScientific:2017qsa,Abbott:2018exr}.}}
\label{jonesy}
\end{center}
\end{figure}

Fig.~\ref{leolausemaus} shows such a $\Lambda_1$-$\Lambda_2$ plot, where the index denotes star number one and star number two. What can be seen is that the plot is symmetric to $\Lambda_1=\Lambda_2$. Along this line the two orbiting stars have the same mass $M= \frac{1}{2}\cdot 2.73~M_{\odot}$.  
Moving towards the origin the star configurations become more compact, because the tidal deformability parameter $\Lambda \propto k_2 \propto C^{-1}$ becomes smaller, see eq.~(\ref{loveandloveismierda}).
The widths of these bands are determined by the small uncertainty in mass $2.73^{+0.04}_{-0.01}M_{\odot}$. 
The lengths of these bands are determined by the uncertainty in mass ratio $1.16 \leq M_{\odot} \leq 1.6$. 
The by far largest values of $\Lambda_1-\Lambda_2$ can be found for a polytrope with $n=1.5$, which is not surprising because of the large radii of these stars, shown in fig.~\ref{ischblasdirdenmarschduarsch} and fig.~\ref{jonesy}. Smaller values in the $\Lambda_1-\Lambda_2$ plot can be found for the $n=1.0$ case, followed by $n=0.7$. These values are shown in the inner right figure on a smaller scale for $\Lambda$ in the $0 \leq \Lambda \leq 20000$. $n=0.7$ is followed by the ultrarelativistic stars with constant speed of sound $c_s^2=1$, the incompressible fluid $n=0.01$ and finally $c_s^2=1/3$. The case for a constant speed of sound EoS $c_s^2=1/3$ exhibits the smallest values of $\Lambda$ by far. These values are a result of the very small radii of the star configurations, which results in rather large value of the compactness, see eq.~(\ref{loveandloveismierda}). The inner upper figure shows the results of $\Lambda$ in the $0 \leq \Lambda \leq 1400$ range. In this range only solutions for the constant speed of sound EoS $c_s^2=1/3$ are found.\\
The continuous black lines in the upper inlaid figure in figure \ref{leolausemaus} denote the 90\% and 50\% credibility level by LIGO \cite{TheLIGOScientific:2017qsa}.  
These lines correspond to a probability density. It is to 90\% propable that a star's tidal deformability parameter $\Lambda$ lies below that line. 50\% of the probability density is located at even lower values.
Every result apart from the $c_s^2=1/3$ lies well outside of the 90\% credibility region, see upper inlaid figure in fig.~\ref{leolausemaus}.

Figure \ref{jonesy} shows the radius and the mass as a function of the compactness for the polytropes and the constant speed of sound EoSs. 
The radii in the upper figure in fig.~\ref{jonesy} are those of stars in a mass range $1.1 \leq M/M_{\odot} \leq 1.6$. The mass as a function of the compactness is shown in the lower figure. Stars in this range can collide to form a total mass of the proto neutron star $2.73^{+0.04}_{-0.01}M_{\odot}$ according to \cite{TheLIGOScientific:2017qsa,Abbott:2018exr}.
It can be seen that the bands in the $\Lambda$-$\Lambda$ plot in fig.~\ref{leolausemaus} develop as follows: The largest values of $\Lambda$ are found for a polytropic index $n=1.5$ $\rightarrow$ $n=1.0$ $\rightarrow$ $n=0.7$ $\rightarrow$ $c_s^2=1$ $\rightarrow$ $n=0.01$ $\rightarrow$ $c_s^2=1/3$. The same development is found in fig.~\ref{jonesy}, i.e. a smaller radius at a given mass yields smaller values of the tidal deformability parameter $\Lambda$. The development of the Love number $k_2$ in fig.~\ref{sostellstdudiraankidmnappingvor} does not follow this development, which means that the compactness gives the main contribution to the tidal deformability parameter $\Lambda$.

The more compact a compact star is, the more the tidal deformability parameter $\Lambda$ is influenced by the compactness. The influence of $k_2$ decreases with increasing compactness. In the Newtonian limit $C \rightarrow 0$ and $\Lambda$ is determined solely via $k_2$.
Although the compactness gives the main contribution in eq.~(\ref{loveusorhateus}), 
$k_2(y,C)$ must definitely not be neglected. 
The actual value of a 1.4$M_{\odot}$ star has to be in a range  $\Lambda=300^{+420}_{-230}$ \cite{TheLIGOScientific:2017qsa,Abbott:2018exr}. 
This is a result of the deformation of spacetime in GR due to terms $\propto M/R$, see eq.~(\ref{nindalf_fennfeld}) in sec.~\ref{mannomausinger}. 


\section{Conclusions and Outlook}
The intention of this article is to provide undergraduate students or senior physics students a simple and pedagogical introduction to a very modern topic in stellar physics, namely, the deformation that takes place in a star that is rotating or which is subject to a tidal field. \\  
We have shown that the rotation of a star around an arbitrary axis obeys the same physics as the application of an external tidal field. Following the article by A.E.H. Love from 1906 \cite{Love:1908tua} our analysis aimed at the quantity called Love number, $k_2$. It is shown how the Love number $k_2$ is related to the quadrupole parameter $J_2$, which is a result of the evaluation of the quadrupole tensor $Q_{ij}$ in classical physics. The tidal Love number $k_2$ is furthermore related
to the tidal deformability parameter $\Lambda $, which, since 2017, is subject in almost every publication concerning compact stars \cite{Hinderer:2007mb,Flanagan:2007ix,Hinderer:2009ca,Postnikov:2010yn}. This is due to the detection of a gravitational wave signal from a binary neutron star merger \cite{TheLIGOScientific:2017qsa,Abbott:2018wiz,Annala:2017llu}. Our main intention in this article is to detangle these quantities and to explain their features vividly.\\ 
We find that the classical derivation of the formalism concerning tidal diruptions yields the correct limit taken from general
relativity, which is adressed after the classical analysis. 
Polytropic equations of state (EoSs) and constant speed of sound EoSs have been introduced as useful examples that yield simplified solutions. 
Two analytical solutions for the tidal Love number $k_2$ are derived and evaluated. The physical implications give direction for a microscopic treatment and are therefore discussed in detail.
For an incompressible fluid the polytropic index is $n = 0$. Classically the compactness $C$ does not enter, hence $k_2$ depends only on a perturbative quantity called $y$. This limiting case gives
$k_2(y=-1)=0.75$. In general relativity $k_2(y,C)$ decreases with increasing compactness.
In the case $n=1$ the analysis is also given in some detail. 
This is suitable for learning certain techniques for handling differential equations which are rarely explained in textbooks and to get acquainted with Bessel functions. 
By solving the TOV equations it is shown that the mass radius relations scale with a constant quantity related to the
EoS. The scaling makes the compactness $C$ and the tidal Love number $k_2$ independent on the constant quantity from the EoSs. The larger the polytropic index n in the EoSs, the smaller is $k_2$. 
The same statement is valid for constant speed of sound EoSs. 
Classically the tidal deformability is described only by the Love number $k_2$. In general relativity the compactness is relevant  and gives
the main contribution to the tidal deformability parameter $\Lambda $, although k$_2$ is not negligible. This is a result of the deformation of spacetime in GR due to terms $\propto $ $M/R$ within the relativistic equations.\\
At the end of 2019 the NASA NICER mission \cite{Raaijmakers:2019qny} investigated the mass and the radius of the millisecond pulsar PSR J0030+0451. This measurement is a first step to infer central densities of compact stars and so to constrain the underlying EoS even more. Together with other GW events which are expected in the future \cite{Abbott:2018qee,Abbott:2020khf}, this article hopefully contributed to a better understanding of the quantities related to the tidal deformability of compact stars and the corresponding relation to classical physics.
\begin{acknowledgments}
The author would like to thank J\"urgen Schaffner-Bielich for helpful comments, reading of the draft and also for valuable discussions on this project. In particular the author would also like to thank Alfred Ziegler and Luciano Rezzolla for careful reading of the drafts, hints and explanations to the equations. 
Michael Wondrak, Yannick Dengler and Felix Ahlbrecht read the draft and made suggestions to its comprehensibility.
\end{acknowledgments}
\bibliography{neue_bib}

\begin{thebibliography}{91}%
\makeatletter
\providecommand \@ifxundefined [1]{%
 \@ifx{#1\undefined}
}%
\providecommand \@ifnum [1]{%
 \ifnum #1\expandafter \@firstoftwo
 \else \expandafter \@secondoftwo
 \fi
}%
\providecommand \@ifx [1]{%
 \ifx #1\expandafter \@firstoftwo
 \else \expandafter \@secondoftwo
 \fi
}%
\providecommand \natexlab [1]{#1}%
\providecommand \enquote  [1]{``#1''}%
\providecommand \bibnamefont  [1]{#1}%
\providecommand \bibfnamefont [1]{#1}%
\providecommand \citenamefont [1]{#1}%
\providecommand \href@noop [0]{\@secondoftwo}%
\providecommand \href [0]{\begingroup \@sanitize@url \@href}%
\providecommand \@href[1]{\@@startlink{#1}\@@href}%
\providecommand \@@href[1]{\endgroup#1\@@endlink}%
\providecommand \@sanitize@url [0]{\catcode `\\12\catcode `\$12\catcode
  `\&12\catcode `\#12\catcode `\^12\catcode `\_12\catcode `\%12\relax}%
\providecommand \@@startlink[1]{}%
\providecommand \@@endlink[0]{}%
\providecommand \url  [0]{\begingroup\@sanitize@url \@url }%
\providecommand \@url [1]{\endgroup\@href {#1}{\urlprefix }}%
\providecommand \urlprefix  [0]{URL }%
\providecommand \Eprint [0]{\href }%
\providecommand \doibase [0]{http://dx.doi.org/}%
\providecommand \selectlanguage [0]{\@gobble}%
\providecommand \bibinfo  [0]{\@secondoftwo}%
\providecommand \bibfield  [0]{\@secondoftwo}%
\providecommand \translation [1]{[#1]}%
\providecommand \BibitemOpen [0]{}%
\providecommand \bibitemStop [0]{}%
\providecommand \bibitemNoStop [0]{.\EOS\space}%
\providecommand \EOS [0]{\spacefactor3000\relax}%
\providecommand \BibitemShut  [1]{\csname bibitem#1\endcsname}%
\let\auto@bib@innerbib\@empty
\bibitem [{\citenamefont {Abbott}\ \emph
  {et~al.}(2016{\natexlab{a}})\citenamefont {Abbott} \emph
  {et~al.}}]{Abbott:2016blz}%
  \BibitemOpen
  \bibfield  {author} {\bibinfo {author} {\bibfnamefont {B.~P.}\ \bibnamefont
  {Abbott}} \emph {et~al.} (\bibinfo {collaboration} {LIGO Scientific,
  Virgo}),\ }\href {\doibase 10.1103/PhysRevLett.116.061102} {\bibfield
  {journal} {\bibinfo  {journal} {Phys. Rev. Lett.}\ }\textbf {\bibinfo
  {volume} {116}},\ \bibinfo {pages} {061102} (\bibinfo {year}
  {2016}{\natexlab{a}})},\ \Eprint {http://arxiv.org/abs/1602.03837}
  {arXiv:1602.03837 [gr-qc]} \BibitemShut {NoStop}%
\bibitem [{\citenamefont {Abbott}\ \emph
  {et~al.}(2016{\natexlab{b}})\citenamefont {Abbott} \emph
  {et~al.}}]{Abbott:2016nmj}%
  \BibitemOpen
  \bibfield  {author} {\bibinfo {author} {\bibfnamefont {B.~P.}\ \bibnamefont
  {Abbott}} \emph {et~al.} (\bibinfo {collaboration} {LIGO Scientific,
  Virgo}),\ }\href {\doibase 10.1103/PhysRevLett.116.241103} {\bibfield
  {journal} {\bibinfo  {journal} {Phys. Rev. Lett.}\ }\textbf {\bibinfo
  {volume} {116}},\ \bibinfo {pages} {241103} (\bibinfo {year}
  {2016}{\natexlab{b}})},\ \Eprint {http://arxiv.org/abs/1606.04855}
  {arXiv:1606.04855 [gr-qc]} \BibitemShut {NoStop}%
\bibitem [{\citenamefont {Abbott}\ \emph
  {et~al.}(2016{\natexlab{c}})\citenamefont {Abbott} \emph
  {et~al.}}]{TheLIGOScientific:2016pea}%
  \BibitemOpen
  \bibfield  {author} {\bibinfo {author} {\bibfnamefont {B.~P.}\ \bibnamefont
  {Abbott}} \emph {et~al.} (\bibinfo {collaboration} {LIGO Scientific,
  Virgo}),\ }\href {\doibase 10.1103/PhysRevX.6.041015,
  10.1103/PhysRevX.8.039903} {\bibfield  {journal} {\bibinfo  {journal} {Phys.
  Rev.}\ }\textbf {\bibinfo {volume} {X6}},\ \bibinfo {pages} {041015}
  (\bibinfo {year} {2016}{\natexlab{c}})},\ \bibinfo {note} {[erratum: Phys.
  Rev.X8,no.3,039903(2018)]},\ \Eprint {http://arxiv.org/abs/1606.04856}
  {arXiv:1606.04856 [gr-qc]} \BibitemShut {NoStop}%
\bibitem [{\citenamefont {Abbott}\ \emph {et~al.}(2017)\citenamefont {Abbott}
  \emph {et~al.}}]{TheLIGOScientific:2017qsa}%
  \BibitemOpen
  \bibfield  {author} {\bibinfo {author} {\bibfnamefont {B.}~\bibnamefont
  {Abbott}} \emph {et~al.} (\bibinfo {collaboration} {Virgo, LIGO
  Scientific}),\ }\href {\doibase 10.1103/PhysRevLett.119.161101} {\bibfield
  {journal} {\bibinfo  {journal} {Phys. Rev. Lett.}\ }\textbf {\bibinfo
  {volume} {119}},\ \bibinfo {pages} {161101} (\bibinfo {year} {2017})},\
  \Eprint {http://arxiv.org/abs/1710.05832} {arXiv:1710.05832 [gr-qc]}
  \BibitemShut {NoStop}%
\bibitem [{\citenamefont {Abbott}\ \emph
  {et~al.}(2019{\natexlab{a}})\citenamefont {Abbott} \emph
  {et~al.}}]{Abbott:2018wiz}%
  \BibitemOpen
  \bibfield  {author} {\bibinfo {author} {\bibfnamefont {B.~P.}\ \bibnamefont
  {Abbott}} \emph {et~al.} (\bibinfo {collaboration} {LIGO Scientific,
  Virgo}),\ }\href {\doibase 10.1103/PhysRevX.9.011001} {\bibfield  {journal}
  {\bibinfo  {journal} {Phys. Rev.}\ }\textbf {\bibinfo {volume} {X9}},\
  \bibinfo {pages} {011001} (\bibinfo {year} {2019}{\natexlab{a}})},\ \Eprint
  {http://arxiv.org/abs/1805.11579} {arXiv:1805.11579 [gr-qc]} \BibitemShut
  {NoStop}%
\bibitem [{\citenamefont {Annala}\ \emph {et~al.}(2018)\citenamefont {Annala},
  \citenamefont {Gorda}, \citenamefont {Kurkela},\ and\ \citenamefont
  {Vuorinen}}]{Annala:2017llu}%
  \BibitemOpen
  \bibfield  {author} {\bibinfo {author} {\bibfnamefont {E.}~\bibnamefont
  {Annala}}, \bibinfo {author} {\bibfnamefont {T.}~\bibnamefont {Gorda}},
  \bibinfo {author} {\bibfnamefont {A.}~\bibnamefont {Kurkela}}, \ and\
  \bibinfo {author} {\bibfnamefont {A.}~\bibnamefont {Vuorinen}},\ }\href
  {\doibase 10.1103/PhysRevLett.120.172703} {\bibfield  {journal} {\bibinfo
  {journal} {Phys. Rev. Lett.}\ }\textbf {\bibinfo {volume} {120}},\ \bibinfo
  {pages} {172703} (\bibinfo {year} {2018})},\ \Eprint
  {http://arxiv.org/abs/1711.02644} {arXiv:1711.02644 [astro-ph.HE]}
  \BibitemShut {NoStop}%
\bibitem [{\citenamefont {Hinderer}(2008)}]{Hinderer:2007mb}%
  \BibitemOpen
  \bibfield  {author} {\bibinfo {author} {\bibfnamefont {T.}~\bibnamefont
  {Hinderer}},\ }\href {\doibase 10.1086/533487} {\bibfield  {journal}
  {\bibinfo  {journal} {Astrophys. J.}\ }\textbf {\bibinfo {volume} {677}},\
  \bibinfo {pages} {1216} (\bibinfo {year} {2008})},\ \Eprint
  {http://arxiv.org/abs/0711.2420} {arXiv:0711.2420 [astro-ph]} \BibitemShut
  {NoStop}%
\bibitem [{\citenamefont {Flanagan}\ and\ \citenamefont
  {Hinderer}(2008)}]{Flanagan:2007ix}%
  \BibitemOpen
  \bibfield  {author} {\bibinfo {author} {\bibfnamefont {E.~E.}\ \bibnamefont
  {Flanagan}}\ and\ \bibinfo {author} {\bibfnamefont {T.}~\bibnamefont
  {Hinderer}},\ }\href {\doibase 10.1103/PhysRevD.77.021502} {\bibfield
  {journal} {\bibinfo  {journal} {Phys. Rev.}\ }\textbf {\bibinfo {volume}
  {D77}},\ \bibinfo {pages} {021502} (\bibinfo {year} {2008})},\ \Eprint
  {http://arxiv.org/abs/0709.1915} {arXiv:0709.1915 [astro-ph]} \BibitemShut
  {NoStop}%
\bibitem [{\citenamefont {Hinderer}\ \emph {et~al.}(2010)\citenamefont
  {Hinderer}, \citenamefont {Lackey}, \citenamefont {Lang},\ and\ \citenamefont
  {Read}}]{Hinderer:2009ca}%
  \BibitemOpen
  \bibfield  {author} {\bibinfo {author} {\bibfnamefont {T.}~\bibnamefont
  {Hinderer}}, \bibinfo {author} {\bibfnamefont {B.~D.}\ \bibnamefont
  {Lackey}}, \bibinfo {author} {\bibfnamefont {R.~N.}\ \bibnamefont {Lang}}, \
  and\ \bibinfo {author} {\bibfnamefont {J.~S.}\ \bibnamefont {Read}},\ }\href
  {\doibase 10.1103/PhysRevD.81.123016} {\bibfield  {journal} {\bibinfo
  {journal} {Phys. Rev.}\ }\textbf {\bibinfo {volume} {D81}},\ \bibinfo {pages}
  {123016} (\bibinfo {year} {2010})},\ \Eprint {http://arxiv.org/abs/0911.3535}
  {arXiv:0911.3535 [astro-ph.HE]} \BibitemShut {NoStop}%
\bibitem [{\citenamefont {Postnikov}\ \emph {et~al.}(2010)\citenamefont
  {Postnikov}, \citenamefont {Prakash},\ and\ \citenamefont
  {Lattimer}}]{Postnikov:2010yn}%
  \BibitemOpen
  \bibfield  {author} {\bibinfo {author} {\bibfnamefont {S.}~\bibnamefont
  {Postnikov}}, \bibinfo {author} {\bibfnamefont {M.}~\bibnamefont {Prakash}},
  \ and\ \bibinfo {author} {\bibfnamefont {J.~M.}\ \bibnamefont {Lattimer}},\
  }\href {\doibase 10.1103/PhysRevD.82.024016} {\bibfield  {journal} {\bibinfo
  {journal} {Phys. Rev.}\ }\textbf {\bibinfo {volume} {D82}},\ \bibinfo {pages}
  {024016} (\bibinfo {year} {2010})},\ \Eprint {http://arxiv.org/abs/1004.5098}
  {arXiv:1004.5098 [astro-ph.SR]} \BibitemShut {NoStop}%
\bibitem [{\citenamefont {Silbar}\ and\ \citenamefont
  {Reddy}(2004)}]{Silbar04}%
  \BibitemOpen
  \bibfield  {author} {\bibinfo {author} {\bibfnamefont {R.~R.}\ \bibnamefont
  {Silbar}}\ and\ \bibinfo {author} {\bibfnamefont {S.}~\bibnamefont {Reddy}},\
  }\href@noop {} {\bibfield  {journal} {\bibinfo  {journal} {Am. J. Phys.}\
  }\textbf {\bibinfo {volume} {72}},\ \bibinfo {pages} {892} (\bibinfo {year}
  {2004})},\ \Eprint {http://arxiv.org/abs/nucl-th/0309041} {nucl-th/0309041}
  \BibitemShut {NoStop}%
\bibitem [{\citenamefont {Sagert}\ \emph {et~al.}(2006)\citenamefont {Sagert},
  \citenamefont {Hempel}, \citenamefont {Greiner},\ and\ \citenamefont
  {Schaffner-Bielich}}]{Sagert:2005fw}%
  \BibitemOpen
  \bibfield  {author} {\bibinfo {author} {\bibfnamefont {I.}~\bibnamefont
  {Sagert}}, \bibinfo {author} {\bibfnamefont {M.}~\bibnamefont {Hempel}},
  \bibinfo {author} {\bibfnamefont {C.}~\bibnamefont {Greiner}}, \ and\
  \bibinfo {author} {\bibfnamefont {J.}~\bibnamefont {Schaffner-Bielich}},\
  }\href {\doibase 10.1088/0143-0807/27/3/012} {\bibfield  {journal} {\bibinfo
  {journal} {Eur.J.Phys.}\ }\textbf {\bibinfo {volume} {27}},\ \bibinfo {pages}
  {577} (\bibinfo {year} {2006})},\ \Eprint
  {http://arxiv.org/abs/astro-ph/0506417} {arXiv:astro-ph/0506417 [astro-ph]}
  \BibitemShut {NoStop}%
\bibitem [{\citenamefont {Schmitt}(2010)}]{Schmitt:2010pn}%
  \BibitemOpen
  \bibfield  {author} {\bibinfo {author} {\bibfnamefont {A.}~\bibnamefont
  {Schmitt}},\ }\href {\doibase 10.1007/978-3-642-12866-0} {\bibfield
  {journal} {\bibinfo  {journal} {Lect. Notes Phys.}\ }\textbf {\bibinfo
  {volume} {811}},\ \bibinfo {pages} {1} (\bibinfo {year} {2010})},\ \Eprint
  {http://arxiv.org/abs/1001.3294} {arXiv:1001.3294 [astro-ph.SR]} \BibitemShut
  {NoStop}%
\bibitem [{\citenamefont {{Glendenning}}(1997)}]{Glendenning_book}%
  \BibitemOpen
  \bibfield  {author} {\bibinfo {author} {\bibfnamefont {N.~K.}\ \bibnamefont
  {{Glendenning}}},\ }\href@noop {} {\emph {\bibinfo {title} {Compact Stars:
  Nuclear Physics, Particle Physics and General Relativity}}}\ (\bibinfo
  {publisher} {Springer},\ \bibinfo {address} {Berkeley},\ \bibinfo {year}
  {1997})\BibitemShut {NoStop}%
\bibitem [{\citenamefont {Glendenning}(1985)}]{Glendenning:1984jr}%
  \BibitemOpen
  \bibfield  {author} {\bibinfo {author} {\bibfnamefont {N.~K.}\ \bibnamefont
  {Glendenning}},\ }\href@noop {} {\bibfield  {journal} {\bibinfo  {journal}
  {Astrophys. J.}\ }\textbf {\bibinfo {volume} {293}},\ \bibinfo {pages} {470}
  (\bibinfo {year} {1985})}\BibitemShut {NoStop}%
\bibitem [{\citenamefont {Gilson}\ and\ \citenamefont
  {Jaffe}(1993)}]{Gilson93}%
  \BibitemOpen
  \bibfield  {author} {\bibinfo {author} {\bibfnamefont {E.~P.}\ \bibnamefont
  {Gilson}}\ and\ \bibinfo {author} {\bibfnamefont {R.~L.}\ \bibnamefont
  {Jaffe}},\ }\href@noop {} {\bibfield  {journal} {\bibinfo  {journal} {Phys.
  Rev. Lett.}\ }\textbf {\bibinfo {volume} {71}},\ \bibinfo {pages} {332}
  (\bibinfo {year} {1993})}\BibitemShut {NoStop}%
\bibitem [{\citenamefont {Ivanenko}\ and\ \citenamefont
  {Kurdgelaidze}(1965)}]{Ivanenko:1965dg}%
  \BibitemOpen
  \bibfield  {author} {\bibinfo {author} {\bibfnamefont {D.~D.}\ \bibnamefont
  {Ivanenko}}\ and\ \bibinfo {author} {\bibfnamefont {D.~F.}\ \bibnamefont
  {Kurdgelaidze}},\ }\href@noop {} {\bibfield  {journal} {\bibinfo  {journal}
  {Astrophys.}\ }\textbf {\bibinfo {volume} {1}},\ \bibinfo {pages} {251}
  (\bibinfo {year} {1965})}\BibitemShut {NoStop}%
\bibitem [{\citenamefont {Itoh}(1970)}]{Itoh:1970uw}%
  \BibitemOpen
  \bibfield  {author} {\bibinfo {author} {\bibfnamefont {N.}~\bibnamefont
  {Itoh}},\ }\href {\doibase 10.1143/PTP.44.291} {\bibfield  {journal}
  {\bibinfo  {journal} {Prog.Theor.Phys.}\ }\textbf {\bibinfo {volume} {44}},\
  \bibinfo {pages} {291} (\bibinfo {year} {1970})}\BibitemShut {NoStop}%
\bibitem [{\citenamefont {Haensel}\ \emph {et~al.}(1986)\citenamefont
  {Haensel}, \citenamefont {Zdunik},\ and\ \citenamefont
  {Schaeffer}}]{Haensel:1986qb}%
  \BibitemOpen
  \bibfield  {author} {\bibinfo {author} {\bibfnamefont {P.}~\bibnamefont
  {Haensel}}, \bibinfo {author} {\bibfnamefont {J.~L.}\ \bibnamefont {Zdunik}},
  \ and\ \bibinfo {author} {\bibfnamefont {R.}~\bibnamefont {Schaeffer}},\
  }\href@noop {} {\bibfield  {journal} {\bibinfo  {journal} {Astron.
  Astrophys.}\ }\textbf {\bibinfo {volume} {160}},\ \bibinfo {pages} {121}
  (\bibinfo {year} {1986})}\BibitemShut {NoStop}%
\bibitem [{\citenamefont {Olinto}(1987)}]{Olinto87}%
  \BibitemOpen
  \bibfield  {author} {\bibinfo {author} {\bibfnamefont {A.~V.}\ \bibnamefont
  {Olinto}},\ }\href@noop {} {\bibfield  {journal} {\bibinfo  {journal} {Phys.
  Lett. B}\ }\textbf {\bibinfo {volume} {192}},\ \bibinfo {pages} {71}
  (\bibinfo {year} {1987})}\BibitemShut {NoStop}%
\bibitem [{\citenamefont {Bodmer}(1971)}]{Bodmer:1971we}%
  \BibitemOpen
  \bibfield  {author} {\bibinfo {author} {\bibfnamefont {A.~R.}\ \bibnamefont
  {Bodmer}},\ }\href@noop {} {\bibfield  {journal} {\bibinfo  {journal} {Phys.
  Rev. D}\ }\textbf {\bibinfo {volume} {4}},\ \bibinfo {pages} {1601} (\bibinfo
  {year} {1971})}\BibitemShut {NoStop}%
\bibitem [{\citenamefont {Witten}(1984)}]{Witten:1984rs}%
  \BibitemOpen
  \bibfield  {author} {\bibinfo {author} {\bibfnamefont {E.}~\bibnamefont
  {Witten}},\ }\href@noop {} {\bibfield  {journal} {\bibinfo  {journal} {Phys.
  Rev. D}\ }\textbf {\bibinfo {volume} {30}},\ \bibinfo {pages} {272} (\bibinfo
  {year} {1984})}\BibitemShut {NoStop}%
\bibitem [{\citenamefont {Schertler}\ \emph {et~al.}(1999)\citenamefont
  {Schertler}, \citenamefont {Leupold},\ and\ \citenamefont
  {Schaffner-Bielich}}]{Schertler:1999xn}%
  \BibitemOpen
  \bibfield  {author} {\bibinfo {author} {\bibfnamefont {K.}~\bibnamefont
  {Schertler}}, \bibinfo {author} {\bibfnamefont {S.}~\bibnamefont {Leupold}},
  \ and\ \bibinfo {author} {\bibfnamefont {J.}~\bibnamefont
  {Schaffner-Bielich}},\ }\href {\doibase 10.1103/PhysRevC.60.025801}
  {\bibfield  {journal} {\bibinfo  {journal} {Phys.Rev.}\ }\textbf {\bibinfo
  {volume} {C60}},\ \bibinfo {pages} {025801} (\bibinfo {year} {1999})},\
  \Eprint {http://arxiv.org/abs/astro-ph/9901152} {arXiv:astro-ph/9901152
  [astro-ph]} \BibitemShut {NoStop}%
\bibitem [{\citenamefont {Schertler}\ \emph {et~al.}(2000)\citenamefont
  {Schertler}, \citenamefont {Greiner}, \citenamefont {Schaffner-Bielich},\
  and\ \citenamefont {Thoma}}]{Schertler:2000xq}%
  \BibitemOpen
  \bibfield  {author} {\bibinfo {author} {\bibfnamefont {K.}~\bibnamefont
  {Schertler}}, \bibinfo {author} {\bibfnamefont {C.}~\bibnamefont {Greiner}},
  \bibinfo {author} {\bibfnamefont {J.}~\bibnamefont {Schaffner-Bielich}}, \
  and\ \bibinfo {author} {\bibfnamefont {M.~H.}\ \bibnamefont {Thoma}},\ }\href
  {\doibase 10.1016/S0375-9474(00)00305-5} {\bibfield  {journal} {\bibinfo
  {journal} {Nucl. Phys.}\ }\textbf {\bibinfo {volume} {A677}},\ \bibinfo
  {pages} {463} (\bibinfo {year} {2000})},\ \Eprint
  {http://arxiv.org/abs/astro-ph/0001467} {arXiv:astro-ph/0001467 [astro-ph]}
  \BibitemShut {NoStop}%
\bibitem [{\citenamefont {Zacchi}\ \emph {et~al.}(2016)\citenamefont {Zacchi},
  \citenamefont {Hanauske},\ and\ \citenamefont
  {Schaffner-Bielich}}]{Zacchi:2015oma}%
  \BibitemOpen
  \bibfield  {author} {\bibinfo {author} {\bibfnamefont {A.}~\bibnamefont
  {Zacchi}}, \bibinfo {author} {\bibfnamefont {M.}~\bibnamefont {Hanauske}}, \
  and\ \bibinfo {author} {\bibfnamefont {J.}~\bibnamefont
  {Schaffner-Bielich}},\ }\href {\doibase 10.1103/PhysRevD.93.065011}
  {\bibfield  {journal} {\bibinfo  {journal} {Phys. Rev.}\ }\textbf {\bibinfo
  {volume} {D93}},\ \bibinfo {pages} {065011} (\bibinfo {year} {2016})},\
  \Eprint {http://arxiv.org/abs/1510.00180} {arXiv:1510.00180 [nucl-th]}
  \BibitemShut {NoStop}%
\bibitem [{\citenamefont {Zacchi}\ \emph {et~al.}(2017)\citenamefont {Zacchi},
  \citenamefont {Tolos},\ and\ \citenamefont
  {Schaffner-Bielich}}]{Zacchi:2016tjw}%
  \BibitemOpen
  \bibfield  {author} {\bibinfo {author} {\bibfnamefont {A.}~\bibnamefont
  {Zacchi}}, \bibinfo {author} {\bibfnamefont {L.}~\bibnamefont {Tolos}}, \
  and\ \bibinfo {author} {\bibfnamefont {J.}~\bibnamefont
  {Schaffner-Bielich}},\ }\href {\doibase 10.1103/PhysRevD.95.103008}
  {\bibfield  {journal} {\bibinfo  {journal} {Phys. Rev.}\ }\textbf {\bibinfo
  {volume} {D95}},\ \bibinfo {pages} {103008} (\bibinfo {year} {2017})},\
  \Eprint {http://arxiv.org/abs/1612.06167} {arXiv:1612.06167 [astro-ph.HE]}
  \BibitemShut {NoStop}%
\bibitem [{\citenamefont {Zacchi}\ and\ \citenamefont
  {Schaffner-Bielich}(2019)}]{Zacchi:2019ayh}%
  \BibitemOpen
  \bibfield  {author} {\bibinfo {author} {\bibfnamefont {A.}~\bibnamefont
  {Zacchi}}\ and\ \bibinfo {author} {\bibfnamefont {J.}~\bibnamefont
  {Schaffner-Bielich}},\ }\href {\doibase 10.1103/PhysRevD.100.123024}
  {\bibfield  {journal} {\bibinfo  {journal} {Phys. Rev.}\ }\textbf {\bibinfo
  {volume} {D100}},\ \bibinfo {pages} {123024} (\bibinfo {year} {2019})},\
  \Eprint {http://arxiv.org/abs/1909.12071} {arXiv:1909.12071 [nucl-th]}
  \BibitemShut {NoStop}%
\bibitem [{\citenamefont {Chodos}\ \emph {et~al.}(1974)\citenamefont {Chodos},
  \citenamefont {Jaffe}, \citenamefont {Johnson}, \citenamefont {Thorn},\ and\
  \citenamefont {Weisskopf}}]{Chodos74}%
  \BibitemOpen
  \bibfield  {author} {\bibinfo {author} {\bibfnamefont {A.}~\bibnamefont
  {Chodos}}, \bibinfo {author} {\bibfnamefont {R.~L.}\ \bibnamefont {Jaffe}},
  \bibinfo {author} {\bibfnamefont {K.}~\bibnamefont {Johnson}}, \bibinfo
  {author} {\bibfnamefont {C.~B.}\ \bibnamefont {Thorn}}, \ and\ \bibinfo
  {author} {\bibfnamefont {V.~F.}\ \bibnamefont {Weisskopf}},\ }\href@noop {}
  {\bibfield  {journal} {\bibinfo  {journal} {Phys. Rev. D}\ }\textbf {\bibinfo
  {volume} {9}},\ \bibinfo {pages} {3471} (\bibinfo {year} {1974})}\BibitemShut
  {NoStop}%
\bibitem [{\citenamefont {Weissenborn}\ \emph {et~al.}(2011)\citenamefont
  {Weissenborn}, \citenamefont {Sagert}, \citenamefont {Pagliara},
  \citenamefont {Hempel},\ and\ \citenamefont
  {Schaffner-Bielich}}]{Weissenborn:2011qu}%
  \BibitemOpen
  \bibfield  {author} {\bibinfo {author} {\bibfnamefont {S.}~\bibnamefont
  {Weissenborn}}, \bibinfo {author} {\bibfnamefont {I.}~\bibnamefont {Sagert}},
  \bibinfo {author} {\bibfnamefont {G.}~\bibnamefont {Pagliara}}, \bibinfo
  {author} {\bibfnamefont {M.}~\bibnamefont {Hempel}}, \ and\ \bibinfo {author}
  {\bibfnamefont {J.}~\bibnamefont {Schaffner-Bielich}},\ }\href {\doibase
  10.1088/2041-8205/740/1/L14} {\bibfield  {journal} {\bibinfo  {journal}
  {Astrophys.J.}\ }\textbf {\bibinfo {volume} {740}},\ \bibinfo {pages} {L14}
  (\bibinfo {year} {2011})},\ \Eprint {http://arxiv.org/abs/1102.2869}
  {arXiv:1102.2869 [astro-ph.HE]} \BibitemShut {NoStop}%
\bibitem [{\citenamefont {Love}(1908)}]{Love:1908tua}%
  \BibitemOpen
  \bibfield  {author} {\bibinfo {author} {\bibfnamefont {A.~E.~H.}\
  \bibnamefont {Love}},\ }\href@noop {} {\bibfield  {journal} {\bibinfo
  {journal} {Proceedings of the Royal Society of London. Series A, Vol.82,
  Issue 551}\ } (\bibinfo {year} {1908})}\BibitemShut {NoStop}%
\bibitem [{\citenamefont {Kuhn}\ \emph {et~al.}(2012)\citenamefont {Kuhn},
  \citenamefont {Bush}, \citenamefont {Emilio},\ and\ \citenamefont
  {Scholl}}]{Kuhn1638}%
  \BibitemOpen
  \bibfield  {author} {\bibinfo {author} {\bibfnamefont {J.~R.}\ \bibnamefont
  {Kuhn}}, \bibinfo {author} {\bibfnamefont {R.}~\bibnamefont {Bush}}, \bibinfo
  {author} {\bibfnamefont {M.}~\bibnamefont {Emilio}}, \ and\ \bibinfo {author}
  {\bibfnamefont {I.~F.}\ \bibnamefont {Scholl}},\ }\href {\doibase
  10.1126/science.1223231} {\bibfield  {journal} {\bibinfo  {journal}
  {Science}\ }\textbf {\bibinfo {volume} {337}},\ \bibinfo {pages} {1638}
  (\bibinfo {year} {2012})},\ \Eprint
  {http://arxiv.org/abs/http://science.sciencemag.org/content/337/6102/1638.full.pdf}
  {http://science.sciencemag.org/content/337/6102/1638.full.pdf} \BibitemShut
  {NoStop}%
\bibitem [{\citenamefont {{Godier}}\ and\ \citenamefont
  {{Rozelot}}(2001)}]{2001SoPh199217G}%
  \BibitemOpen
  \bibfield  {author} {\bibinfo {author} {\bibfnamefont {S.}~\bibnamefont
  {{Godier}}}\ and\ \bibinfo {author} {\bibfnamefont {J.~P.}\ \bibnamefont
  {{Rozelot}}},\ }\href {\doibase 10.1023/A:1010354901960} {\bibfield
  {journal} {\bibinfo  {journal} {Solar Physics, v. 199, Issue 2, p. 217-229
  (2001)}\ }\textbf {\bibinfo {volume} {199}},\ \bibinfo {pages} {217}
  (\bibinfo {year} {2001})}\BibitemShut {NoStop}%
\bibitem [{\citenamefont {Ragozzine}\ and\ \citenamefont
  {Wolf}(2009)}]{Ragozzine:2008bp}%
  \BibitemOpen
  \bibfield  {author} {\bibinfo {author} {\bibfnamefont {D.}~\bibnamefont
  {Ragozzine}}\ and\ \bibinfo {author} {\bibfnamefont {A.~S.}\ \bibnamefont
  {Wolf}},\ }\href {\doibase 10.1088/0004-637X/698/2/1778} {\bibfield
  {journal} {\bibinfo  {journal} {Astrophys. J.}\ }\textbf {\bibinfo {volume}
  {698}},\ \bibinfo {pages} {1778} (\bibinfo {year} {2009})},\ \Eprint
  {http://arxiv.org/abs/0807.2856} {arXiv:0807.2856 [astro-ph]} \BibitemShut
  {NoStop}%
\bibitem [{\citenamefont {Jordan}\ and\ \citenamefont
  {Bakos}(2008)}]{Jordan:2008zi}%
  \BibitemOpen
  \bibfield  {author} {\bibinfo {author} {\bibfnamefont {A.}~\bibnamefont
  {Jordan}}\ and\ \bibinfo {author} {\bibfnamefont {G.~A.}\ \bibnamefont
  {Bakos}},\ }\href {\doibase 10.1086/590549} {\bibfield  {journal} {\bibinfo
  {journal} {Astrophys. J.}\ }\textbf {\bibinfo {volume} {685}},\ \bibinfo
  {pages} {543} (\bibinfo {year} {2008})},\ \Eprint
  {http://arxiv.org/abs/0806.0630} {arXiv:0806.0630 [astro-ph]} \BibitemShut
  {NoStop}%
\bibitem [{\citenamefont {Lainey}(2016)}]{Lainey_2016}%
  \BibitemOpen
  \bibfield  {author} {\bibinfo {author} {\bibfnamefont {V.}~\bibnamefont
  {Lainey}},\ }\href {\doibase 10.1007/s10569-016-9695-y} {\bibfield  {journal}
  {\bibinfo  {journal} {Celestial Mechanics and Dynamical Astronomy}\ }\textbf
  {\bibinfo {volume} {126}},\ \bibinfo {pages} {145–156} (\bibinfo {year}
  {2016})}\BibitemShut {NoStop}%
\bibitem [{\citenamefont {Iess}\ \emph {et~al.}(2012)\citenamefont {Iess},
  \citenamefont {Jacobson}, \citenamefont {Ducci}, \citenamefont {Stevenson},
  \citenamefont {Lunine}, \citenamefont {Armstrong}, \citenamefont {Asmar},
  \citenamefont {Racioppa}, \citenamefont {Rappaport},\ and\ \citenamefont
  {Tortora}}]{article_titan}%
  \BibitemOpen
  \bibfield  {author} {\bibinfo {author} {\bibfnamefont {L.}~\bibnamefont
  {Iess}}, \bibinfo {author} {\bibfnamefont {R.}~\bibnamefont {Jacobson}},
  \bibinfo {author} {\bibfnamefont {M.}~\bibnamefont {Ducci}}, \bibinfo
  {author} {\bibfnamefont {D.}~\bibnamefont {Stevenson}}, \bibinfo {author}
  {\bibfnamefont {J.}~\bibnamefont {Lunine}}, \bibinfo {author} {\bibfnamefont
  {J.}~\bibnamefont {Armstrong}}, \bibinfo {author} {\bibfnamefont
  {S.}~\bibnamefont {Asmar}}, \bibinfo {author} {\bibfnamefont
  {P.}~\bibnamefont {Racioppa}}, \bibinfo {author} {\bibfnamefont
  {N.}~\bibnamefont {Rappaport}}, \ and\ \bibinfo {author} {\bibfnamefont
  {P.}~\bibnamefont {Tortora}},\ }\href {\doibase 10.1126/science.1219631}
  {\bibfield  {journal} {\bibinfo  {journal} {Science (New York, N.Y.)}\
  }\textbf {\bibinfo {volume} {337}},\ \bibinfo {pages} {457} (\bibinfo {year}
  {2012})}\BibitemShut {NoStop}%
\bibitem [{\citenamefont {Sterne}(1939)}]{10.1093/mnras/99.5.451}%
  \BibitemOpen
  \bibfield  {author} {\bibinfo {author} {\bibfnamefont {T.~E.}\ \bibnamefont
  {Sterne}},\ }\href {\doibase 10.1093/mnras/99.5.451} {\bibfield  {journal}
  {\bibinfo  {journal} {Monthly Notices of the Royal Astronomical Society}\
  }\textbf {\bibinfo {volume} {99}},\ \bibinfo {pages} {451} (\bibinfo {year}
  {1939})},\ \Eprint
  {http://arxiv.org/abs/http://oup.prod.sis.lan/mnras/article-pdf/99/5/451/3286406/mnras99-0451.pdf}
  {http://oup.prod.sis.lan/mnras/article-pdf/99/5/451/3286406/mnras99-0451.pdf}
  \BibitemShut {NoStop}%
\bibitem [{\citenamefont {Wu}\ and\ \citenamefont
  {Goldreich}(2002)}]{Wu:2001wn}%
  \BibitemOpen
  \bibfield  {author} {\bibinfo {author} {\bibfnamefont {Y.}~\bibnamefont
  {Wu}}\ and\ \bibinfo {author} {\bibfnamefont {P.}~\bibnamefont {Goldreich}},\
  }\href {\doibase 10.1086/324193} {\bibfield  {journal} {\bibinfo  {journal}
  {Astrophys. J.}\ }\textbf {\bibinfo {volume} {564}},\ \bibinfo {pages} {1024}
  (\bibinfo {year} {2002})},\ \Eprint {http://arxiv.org/abs/astro-ph/0108499}
  {arXiv:astro-ph/0108499 [astro-ph]} \BibitemShut {NoStop}%
\bibitem [{\citenamefont {{Roxburgh}}(1964)}]{roxburgh}%
  \BibitemOpen
  \bibfield  {author} {\bibinfo {author} {\bibfnamefont {I.}~\bibnamefont
  {{Roxburgh}}},\ }\href {\doibase
  https://www.sciencedirect.com/science/article/pii/0019103564900491?via%3Dihub}
  {\emph {\bibinfo {title} {Solar rotation and the perihelion advance of the
  planets}}},\ Vol.\ \bibinfo {volume} {3 Issue 2 Icarus, p.92-97}\ (\bibinfo
  {year} {1964})\BibitemShut {NoStop}%
\bibitem [{\citenamefont {{{Paterno}, L. and {Sofia}, S. and {di Mauro},
  M.~P.}}(1996)}]{Paterno:1996abc}%
  \BibitemOpen
  \bibfield  {author} {\bibinfo {author} {\bibnamefont {{{Paterno}, L. and
  {Sofia}, S. and {di Mauro}, M.~P.}}},\ }\href@noop {} {\bibfield  {journal}
  {\bibinfo  {journal} {AAP}\ }\textbf {\bibinfo {volume} {314}},\ \bibinfo
  {pages} {p.940} (\bibinfo {year} {1996})}\BibitemShut {NoStop}%
\bibitem [{\citenamefont {Alcock}\ \emph {et~al.}(1986)\citenamefont {Alcock},
  \citenamefont {Farhi},\ and\ \citenamefont {Olinto}}]{Alcock:1986hz}%
  \BibitemOpen
  \bibfield  {author} {\bibinfo {author} {\bibfnamefont {C.}~\bibnamefont
  {Alcock}}, \bibinfo {author} {\bibfnamefont {E.}~\bibnamefont {Farhi}}, \
  and\ \bibinfo {author} {\bibfnamefont {A.}~\bibnamefont {Olinto}},\
  }\href@noop {} {\bibfield  {journal} {\bibinfo  {journal} {Astrophys. J.}\
  }\textbf {\bibinfo {volume} {310}},\ \bibinfo {pages} {261} (\bibinfo {year}
  {1986})}\BibitemShut {NoStop}%
\bibitem [{\citenamefont {Weber}(2005)}]{Weber:2004kj}%
  \BibitemOpen
  \bibfield  {author} {\bibinfo {author} {\bibfnamefont {F.}~\bibnamefont
  {Weber}},\ }\href@noop {} {\bibfield  {journal} {\bibinfo  {journal} {Prog.
  Part. Nucl. Phys.}\ }\textbf {\bibinfo {volume} {54}},\ \bibinfo {pages}
  {193} (\bibinfo {year} {2005})},\ \Eprint
  {http://arxiv.org/abs/astro-ph/0407155} {astro-ph/0407155} \BibitemShut
  {NoStop}%
\bibitem [{\citenamefont {Zacchi}\ \emph {et~al.}(2015)\citenamefont {Zacchi},
  \citenamefont {Stiele},\ and\ \citenamefont
  {Schaffner-Bielich}}]{Zacchi:2015lwa}%
  \BibitemOpen
  \bibfield  {author} {\bibinfo {author} {\bibfnamefont {A.}~\bibnamefont
  {Zacchi}}, \bibinfo {author} {\bibfnamefont {R.}~\bibnamefont {Stiele}}, \
  and\ \bibinfo {author} {\bibfnamefont {J.}~\bibnamefont
  {Schaffner-Bielich}},\ }\href@noop {} {\  (\bibinfo {year} {2015})},\ \Eprint
  {http://arxiv.org/abs/1506.01868} {arXiv:1506.01868 [astro-ph.HE]}
  \BibitemShut {NoStop}%
\bibitem [{\citenamefont {Damour}\ and\ \citenamefont
  {Nagar}(2009)}]{Damour:2009vw}%
  \BibitemOpen
  \bibfield  {author} {\bibinfo {author} {\bibfnamefont {T.}~\bibnamefont
  {Damour}}\ and\ \bibinfo {author} {\bibfnamefont {A.}~\bibnamefont {Nagar}},\
  }\href {\doibase 10.1103/PhysRevD.80.084035} {\bibfield  {journal} {\bibinfo
  {journal} {Phys. Rev.}\ }\textbf {\bibinfo {volume} {D80}},\ \bibinfo {pages}
  {084035} (\bibinfo {year} {2009})},\ \Eprint {http://arxiv.org/abs/0906.0096}
  {arXiv:0906.0096 [gr-qc]} \BibitemShut {NoStop}%
\bibitem [{\citenamefont {Batygin}\ \emph {et~al.}(2009)\citenamefont
  {Batygin}, \citenamefont {Bodenheimer},\ and\ \citenamefont
  {Laughlin}}]{Batygin:2009ps}%
  \BibitemOpen
  \bibfield  {author} {\bibinfo {author} {\bibfnamefont {K.}~\bibnamefont
  {Batygin}}, \bibinfo {author} {\bibfnamefont {P.}~\bibnamefont
  {Bodenheimer}}, \ and\ \bibinfo {author} {\bibfnamefont {G.}~\bibnamefont
  {Laughlin}},\ }\href {\doibase 10.1088/0004-637X/704/1/L49} {\bibfield
  {journal} {\bibinfo  {journal} {Astrophys. J.}\ }\textbf {\bibinfo {volume}
  {704}},\ \bibinfo {pages} {L49} (\bibinfo {year} {2009})},\ \Eprint
  {http://arxiv.org/abs/0907.5019} {arXiv:0907.5019 [astro-ph.EP]} \BibitemShut
  {NoStop}%
\bibitem [{\citenamefont {Becker}\ and\ \citenamefont
  {Batygin}(2013)}]{Becker:2013fma}%
  \BibitemOpen
  \bibfield  {author} {\bibinfo {author} {\bibfnamefont {J.~C.}\ \bibnamefont
  {Becker}}\ and\ \bibinfo {author} {\bibfnamefont {K.}~\bibnamefont
  {Batygin}},\ }\href {\doibase 10.1088/0004-637X/778/2/100} {\bibfield
  {journal} {\bibinfo  {journal} {Astrophys. J.}\ }\textbf {\bibinfo {volume}
  {778}},\ \bibinfo {pages} {100} (\bibinfo {year} {2013})},\ \Eprint
  {http://arxiv.org/abs/1309.5363} {arXiv:1309.5363 [astro-ph.EP]} \BibitemShut
  {NoStop}%
\bibitem [{\citenamefont {Demorest}\ \emph {et~al.}(2010)\citenamefont
  {Demorest}, \citenamefont {Pennucci}, \citenamefont {Ransom}, \citenamefont
  {Roberts},\ and\ \citenamefont {Hessels}}]{Demorest:2010bx}%
  \BibitemOpen
  \bibfield  {author} {\bibinfo {author} {\bibfnamefont {P.}~\bibnamefont
  {Demorest}}, \bibinfo {author} {\bibfnamefont {T.}~\bibnamefont {Pennucci}},
  \bibinfo {author} {\bibfnamefont {S.}~\bibnamefont {Ransom}}, \bibinfo
  {author} {\bibfnamefont {M.}~\bibnamefont {Roberts}}, \ and\ \bibinfo
  {author} {\bibfnamefont {J.}~\bibnamefont {Hessels}},\ }\href {\doibase
  10.1038/nature09466} {\bibfield  {journal} {\bibinfo  {journal} {Nature}\
  }\textbf {\bibinfo {volume} {467}},\ \bibinfo {pages} {1081} (\bibinfo {year}
  {2010})},\ \Eprint {http://arxiv.org/abs/1010.5788} {arXiv:1010.5788
  [astro-ph.HE]} \BibitemShut {NoStop}%
\bibitem [{\citenamefont {Antoniadis}\ \emph {et~al.}(2013)\citenamefont
  {Antoniadis}, \citenamefont {Freire}, \citenamefont {Wex}, \citenamefont
  {Tauris}, \citenamefont {Lynch}, \citenamefont {van Kerkwijk}, \citenamefont
  {Kramer}, \citenamefont {Bassa}, \citenamefont {Dhillon}, \citenamefont
  {Driebe}, \citenamefont {Hessels}, \citenamefont {Kaspi}, \citenamefont
  {Kondratiev}, \citenamefont {Langer}, \citenamefont {Marsh}, \citenamefont
  {McLaughlin}, \citenamefont {Pennucci}, \citenamefont {Ransom}, \citenamefont
  {Stairs}, \citenamefont {van Leeuwen}, \citenamefont {Verbiest},\ and\
  \citenamefont {Whelan}}]{Antoniadis:2013pzd}%
  \BibitemOpen
  \bibfield  {author} {\bibinfo {author} {\bibfnamefont {J.}~\bibnamefont
  {Antoniadis}}, \bibinfo {author} {\bibfnamefont {P.~C.}\ \bibnamefont
  {Freire}}, \bibinfo {author} {\bibfnamefont {N.}~\bibnamefont {Wex}},
  \bibinfo {author} {\bibfnamefont {T.~M.}\ \bibnamefont {Tauris}}, \bibinfo
  {author} {\bibfnamefont {R.~S.}\ \bibnamefont {Lynch}}, \bibinfo {author}
  {\bibfnamefont {M.~H.}\ \bibnamefont {van Kerkwijk}}, \bibinfo {author}
  {\bibfnamefont {M.}~\bibnamefont {Kramer}}, \bibinfo {author} {\bibfnamefont
  {C.}~\bibnamefont {Bassa}}, \bibinfo {author} {\bibfnamefont {V.~S.}\
  \bibnamefont {Dhillon}}, \bibinfo {author} {\bibfnamefont {T.}~\bibnamefont
  {Driebe}}, \bibinfo {author} {\bibfnamefont {J.~W.~T.}\ \bibnamefont
  {Hessels}}, \bibinfo {author} {\bibfnamefont {V.~M.}\ \bibnamefont {Kaspi}},
  \bibinfo {author} {\bibfnamefont {V.~I.}\ \bibnamefont {Kondratiev}},
  \bibinfo {author} {\bibfnamefont {N.}~\bibnamefont {Langer}}, \bibinfo
  {author} {\bibfnamefont {T.~R.}\ \bibnamefont {Marsh}}, \bibinfo {author}
  {\bibfnamefont {M.~A.}\ \bibnamefont {McLaughlin}}, \bibinfo {author}
  {\bibfnamefont {T.~T.}\ \bibnamefont {Pennucci}}, \bibinfo {author}
  {\bibfnamefont {S.~M.}\ \bibnamefont {Ransom}}, \bibinfo {author}
  {\bibfnamefont {I.~H.}\ \bibnamefont {Stairs}}, \bibinfo {author}
  {\bibfnamefont {J.}~\bibnamefont {van Leeuwen}}, \bibinfo {author}
  {\bibfnamefont {J.~P.~W.}\ \bibnamefont {Verbiest}}, \ and\ \bibinfo {author}
  {\bibfnamefont {D.~G.}\ \bibnamefont {Whelan}},\ }\href {\doibase
  10.1126/science.1233232} {\bibfield  {journal} {\bibinfo  {journal}
  {Science}\ }\textbf {\bibinfo {volume} {340}},\ \bibinfo {pages} {6131}
  (\bibinfo {year} {2013})},\ \Eprint {http://arxiv.org/abs/1304.6875}
  {arXiv:1304.6875 [astro-ph.HE]} \BibitemShut {NoStop}%
\bibitem [{\citenamefont {Fonseca}\ \emph {et~al.}(2016)\citenamefont {Fonseca}
  \emph {et~al.}}]{Fonseca:2016tux}%
  \BibitemOpen
  \bibfield  {author} {\bibinfo {author} {\bibfnamefont {E.}~\bibnamefont
  {Fonseca}} \emph {et~al.},\ }\href {\doibase 10.3847/0004-637X/832/2/167}
  {\bibfield  {journal} {\bibinfo  {journal} {Astrophys. J.}\ }\textbf
  {\bibinfo {volume} {832}},\ \bibinfo {pages} {167} (\bibinfo {year}
  {2016})},\ \Eprint {http://arxiv.org/abs/1603.00545} {arXiv:1603.00545
  [astro-ph.HE]} \BibitemShut {NoStop}%
\bibitem [{\citenamefont {Cromartie}\ \emph {et~al.}(2019)\citenamefont
  {Cromartie} \emph {et~al.}}]{Cromartie:2019kug}%
  \BibitemOpen
  \bibfield  {author} {\bibinfo {author} {\bibfnamefont {H.~T.}\ \bibnamefont
  {Cromartie}} \emph {et~al.},\ }\href@noop {} {\  (\bibinfo {year} {2019})},\
  \Eprint {http://arxiv.org/abs/1904.06759} {arXiv:1904.06759 [astro-ph.HE]}
  \BibitemShut {NoStop}%
\bibitem [{\citenamefont {K{\"a}mpfer}(1981{\natexlab{a}})}]{Kaempfer:1981a}%
  \BibitemOpen
  \bibfield  {author} {\bibinfo {author} {\bibfnamefont {B.}~\bibnamefont
  {K{\"a}mpfer}},\ }\href@noop {} {\bibfield  {journal} {\bibinfo  {journal}
  {Phys. Lett.}\ }\textbf {\bibinfo {volume} {101B}},\ \bibinfo {pages} {366}
  (\bibinfo {year} {1981}{\natexlab{a}})}\BibitemShut {NoStop}%
\bibitem [{\citenamefont {K{\"a}mpfer}(1981{\natexlab{b}})}]{Kaempfer81}%
  \BibitemOpen
  \bibfield  {author} {\bibinfo {author} {\bibfnamefont {B.}~\bibnamefont
  {K{\"a}mpfer}},\ }\href@noop {} {\bibfield  {journal} {\bibinfo  {journal}
  {J. Phys. A}\ }\textbf {\bibinfo {volume} {14}},\ \bibinfo {pages} {L471}
  (\bibinfo {year} {1981}{\natexlab{b}})}\BibitemShut {NoStop}%
\bibitem [{\citenamefont {K{\"a}mpfer}(1982)}]{Kaempfer82}%
  \BibitemOpen
  \bibfield  {author} {\bibinfo {author} {\bibfnamefont {B.}~\bibnamefont
  {K{\"a}mpfer}},\ }\href@noop {} {\bibfield  {journal} {\bibinfo  {journal}
  {Astron. Nachr.}\ }\textbf {\bibinfo {volume} {303}},\ \bibinfo {pages} {231}
  (\bibinfo {year} {1982})}\BibitemShut {NoStop}%
\bibitem [{\citenamefont {K{\"a}mpfer}(1983{\natexlab{a}})}]{Kaempfer83a}%
  \BibitemOpen
  \bibfield  {author} {\bibinfo {author} {\bibfnamefont {B.}~\bibnamefont
  {K{\"a}mpfer}},\ }\href@noop {} {\bibfield  {journal} {\bibinfo  {journal}
  {J. Phys. A}\ }\textbf {\bibinfo {volume} {16}},\ \bibinfo {pages} {633}
  (\bibinfo {year} {1983}{\natexlab{a}})}\BibitemShut {NoStop}%
\bibitem [{\citenamefont {K{\"a}mpfer}(1983{\natexlab{b}})}]{Kaempfer83b}%
  \BibitemOpen
  \bibfield  {author} {\bibinfo {author} {\bibfnamefont {B.}~\bibnamefont
  {K{\"a}mpfer}},\ }\href@noop {} {\bibfield  {journal} {\bibinfo  {journal}
  {Astron. Nachr.}\ }\textbf {\bibinfo {volume} {304}},\ \bibinfo {pages} {167}
  (\bibinfo {year} {1983}{\natexlab{b}})}\BibitemShut {NoStop}%
\bibitem [{\citenamefont {K{\"a}mpfer}(1985)}]{Kaempfer85}%
  \BibitemOpen
  \bibfield  {author} {\bibinfo {author} {\bibfnamefont {B.}~\bibnamefont
  {K{\"a}mpfer}},\ }\href@noop {} {\bibfield  {journal} {\bibinfo  {journal}
  {Phys. Lett.}\ }\textbf {\bibinfo {volume} {153B}},\ \bibinfo {pages} {121}
  (\bibinfo {year} {1985})}\BibitemShut {NoStop}%
\bibitem [{\citenamefont {Glendenning}\ and\ \citenamefont
  {Kettner}(2000)}]{Glendenning:1998ag}%
  \BibitemOpen
  \bibfield  {author} {\bibinfo {author} {\bibfnamefont {N.~K.}\ \bibnamefont
  {Glendenning}}\ and\ \bibinfo {author} {\bibfnamefont {C.}~\bibnamefont
  {Kettner}},\ }\href@noop {} {\bibfield  {journal} {\bibinfo  {journal}
  {Astron. Astrophys.}\ }\textbf {\bibinfo {volume} {353}},\ \bibinfo {pages}
  {L9} (\bibinfo {year} {2000})},\ \Eprint
  {http://arxiv.org/abs/astro-ph/9807155} {astro-ph/9807155} \BibitemShut
  {NoStop}%
\bibitem [{\citenamefont {Schaeffer}\ and\ \citenamefont
  {Haensel}(1983)}]{zdunikhaenselschaeffer:1983pti}%
  \BibitemOpen
  \bibfield  {author} {\bibinfo {author} {\bibfnamefont {Z.~L.}\ \bibnamefont
  {Schaeffer}, \bibfnamefont {R.}}\ and\ \bibinfo {author} {\bibfnamefont
  {P.}~\bibnamefont {Haensel}},\ }\href@noop {} {\bibfield  {journal} {\bibinfo
   {journal} {Astron. Astrophysics, 126 (121-145)}\ } (\bibinfo {year}
  {1983})}\BibitemShut {NoStop}%
\bibitem [{\citenamefont {Sagert}\ \emph {et~al.}(2008)\citenamefont {Sagert},
  \citenamefont {Pagliara}, \citenamefont {Hempel},\ and\ \citenamefont
  {Schaffner-Bielich}}]{Sagert:2008uq}%
  \BibitemOpen
  \bibfield  {author} {\bibinfo {author} {\bibfnamefont {I.}~\bibnamefont
  {Sagert}}, \bibinfo {author} {\bibfnamefont {G.}~\bibnamefont {Pagliara}},
  \bibinfo {author} {\bibfnamefont {M.}~\bibnamefont {Hempel}}, \ and\ \bibinfo
  {author} {\bibfnamefont {J.}~\bibnamefont {Schaffner-Bielich}},\ }\href
  {\doibase 10.1088/0954-3899/35/10/104079} {\bibfield  {journal} {\bibinfo
  {journal} {J.Phys.}\ }\textbf {\bibinfo {volume} {G35}},\ \bibinfo {pages}
  {104079} (\bibinfo {year} {2008})},\ \Eprint {http://arxiv.org/abs/0808.1049}
  {arXiv:0808.1049 [astro-ph]} \BibitemShut {NoStop}%
\bibitem [{\citenamefont {Schaffner-Bielich}(2010)}]{SchaffnerBielich:2010am}%
  \BibitemOpen
  \bibfield  {author} {\bibinfo {author} {\bibfnamefont {J.}~\bibnamefont
  {Schaffner-Bielich}},\ }\href {\doibase 10.1016/j.nuclphysa.2010.01.203}
  {\bibfield  {journal} {\bibinfo  {journal} {Nucl.Phys.}\ }\textbf {\bibinfo
  {volume} {A835}},\ \bibinfo {pages} {279} (\bibinfo {year} {2010})},\ \Eprint
  {http://arxiv.org/abs/1002.1658} {arXiv:1002.1658 [nucl-th]} \BibitemShut
  {NoStop}%
\bibitem [{\citenamefont {Weissenborn}\ \emph {et~al.}(2012)\citenamefont
  {Weissenborn}, \citenamefont {Chatterjee},\ and\ \citenamefont
  {Schaffner-Bielich}}]{Weissenborn:2011ut}%
  \BibitemOpen
  \bibfield  {author} {\bibinfo {author} {\bibfnamefont {S.}~\bibnamefont
  {Weissenborn}}, \bibinfo {author} {\bibfnamefont {D.}~\bibnamefont
  {Chatterjee}}, \ and\ \bibinfo {author} {\bibfnamefont {J.}~\bibnamefont
  {Schaffner-Bielich}},\ }\href {\doibase 10.1103/PhysRevC.85.065802}
  {\bibfield  {journal} {\bibinfo  {journal} {Phys.Rev.}\ }\textbf {\bibinfo
  {volume} {C85}},\ \bibinfo {pages} {065802} (\bibinfo {year} {2012})},\
  \Eprint {http://arxiv.org/abs/1112.0234} {arXiv:1112.0234 [astro-ph.HE]}
  \BibitemShut {NoStop}%
\bibitem [{\citenamefont {Alford}\ \emph {et~al.}(2014)\citenamefont {Alford},
  \citenamefont {Han},\ and\ \citenamefont {Prakash}}]{Alford:2014dva}%
  \BibitemOpen
  \bibfield  {author} {\bibinfo {author} {\bibfnamefont {M.~G.}\ \bibnamefont
  {Alford}}, \bibinfo {author} {\bibfnamefont {S.}~\bibnamefont {Han}}, \ and\
  \bibinfo {author} {\bibfnamefont {M.}~\bibnamefont {Prakash}},\ }\href
  {\doibase 10.7566/JPSCP.1.013041} {\bibfield  {journal} {\bibinfo  {journal}
  {JPS Conf.Proc.}\ }\textbf {\bibinfo {volume} {1}},\ \bibinfo {pages}
  {013041} (\bibinfo {year} {2014})}\BibitemShut {NoStop}%
\bibitem [{\citenamefont {Alford}\ \emph {et~al.}(2015)\citenamefont {Alford},
  \citenamefont {Burgio}, \citenamefont {Han}, \citenamefont {Taranto},\ and\
  \citenamefont {Zappalà}}]{Alford:2015dpa}%
  \BibitemOpen
  \bibfield  {author} {\bibinfo {author} {\bibfnamefont {M.~G.}\ \bibnamefont
  {Alford}}, \bibinfo {author} {\bibfnamefont {G.}~\bibnamefont {Burgio}},
  \bibinfo {author} {\bibfnamefont {S.}~\bibnamefont {Han}}, \bibinfo {author}
  {\bibfnamefont {G.}~\bibnamefont {Taranto}}, \ and\ \bibinfo {author}
  {\bibfnamefont {D.}~\bibnamefont {Zappalà}},\ }\href@noop {} {\  (\bibinfo
  {year} {2015})},\ \Eprint {http://arxiv.org/abs/1501.07902} {arXiv:1501.07902
  [nucl-th]} \BibitemShut {NoStop}%
\bibitem [{\citenamefont {Benic}\ \emph {et~al.}(2015)\citenamefont {Benic},
  \citenamefont {Blaschke}, \citenamefont {Alvarez-Castillo}, \citenamefont
  {Fischer},\ and\ \citenamefont {Typel}}]{Benic:2014jia}%
  \BibitemOpen
  \bibfield  {author} {\bibinfo {author} {\bibfnamefont {S.}~\bibnamefont
  {Benic}}, \bibinfo {author} {\bibfnamefont {D.}~\bibnamefont {Blaschke}},
  \bibinfo {author} {\bibfnamefont {D.~E.}\ \bibnamefont {Alvarez-Castillo}},
  \bibinfo {author} {\bibfnamefont {T.}~\bibnamefont {Fischer}}, \ and\
  \bibinfo {author} {\bibfnamefont {S.}~\bibnamefont {Typel}},\ }\href
  {\doibase 10.1051/0004-6361/201425318} {\bibfield  {journal} {\bibinfo
  {journal} {Astron.Astrophys.}\ }\textbf {\bibinfo {volume} {577}},\ \bibinfo
  {pages} {A40} (\bibinfo {year} {2015})},\ \Eprint
  {http://arxiv.org/abs/1411.2856} {arXiv:1411.2856 [astro-ph.HE]} \BibitemShut
  {NoStop}%
\bibitem [{\citenamefont {Blaschke}\ and\ \citenamefont
  {Alvarez-Castillo}(2015)}]{Blaschke:2015uva}%
  \BibitemOpen
  \bibfield  {author} {\bibinfo {author} {\bibfnamefont {D.}~\bibnamefont
  {Blaschke}}\ and\ \bibinfo {author} {\bibfnamefont {D.~E.}\ \bibnamefont
  {Alvarez-Castillo}},\ }\href@noop {} {\  (\bibinfo {year} {2015})},\ \Eprint
  {http://arxiv.org/abs/1503.03834} {arXiv:1503.03834 [astro-ph.HE]}
  \BibitemShut {NoStop}%
\bibitem [{\citenamefont {Christian}\ \emph {et~al.}(2018)\citenamefont
  {Christian}, \citenamefont {Zacchi},\ and\ \citenamefont
  {Schaffner-Bielich}}]{Christian:2017jni}%
  \BibitemOpen
  \bibfield  {author} {\bibinfo {author} {\bibfnamefont {J.-E.}\ \bibnamefont
  {Christian}}, \bibinfo {author} {\bibfnamefont {A.}~\bibnamefont {Zacchi}}, \
  and\ \bibinfo {author} {\bibfnamefont {J.}~\bibnamefont
  {Schaffner-Bielich}},\ }\href {\doibase 10.1140/epja/i2018-12472-y}
  {\bibfield  {journal} {\bibinfo  {journal} {Eur. Phys. J.}\ }\textbf
  {\bibinfo {volume} {A54}},\ \bibinfo {pages} {28} (\bibinfo {year} {2018})},\
  \Eprint {http://arxiv.org/abs/1707.07524} {arXiv:1707.07524 [astro-ph.HE]}
  \BibitemShut {NoStop}%
\bibitem [{\citenamefont {{Zel'dovich}}\ and\ \citenamefont
  {{Novikov}}(1971)}]{Zeldovich_book}%
  \BibitemOpen
  \bibfield  {author} {\bibinfo {author} {\bibfnamefont {Y.~B.}\ \bibnamefont
  {{Zel'dovich}}}\ and\ \bibinfo {author} {\bibfnamefont {I.~D.}\ \bibnamefont
  {{Novikov}}},\ }\href@noop {} {\emph {\bibinfo {title} {Relativistic
  astrophysics. Volume 1: Stars and relativity}}}\ (\bibinfo  {publisher}
  {University of Chicago Press},\ \bibinfo {address} {Chicago},\ \bibinfo
  {year} {1971})\BibitemShut {NoStop}%
\bibitem [{\citenamefont {{Fliessbach}}(2003)}]{fliessbach_book}%
  \BibitemOpen
  \bibfield  {author} {\bibinfo {author} {\bibfnamefont {T.}~\bibnamefont
  {{Fliessbach}}},\ }\href@noop {} {\emph {\bibinfo {title} {Allgemeine
  Relativit\"atstheorie}}}\ (\bibinfo  {publisher} {Spektrum},\ \bibinfo
  {address} {Siegen},\ \bibinfo {year} {2003})\BibitemShut {NoStop}%
\bibitem [{\citenamefont {{Stannard}}(2008)}]{Stannard_book}%
  \BibitemOpen
  \bibfield  {author} {\bibinfo {author} {\bibfnamefont {R.}~\bibnamefont
  {{Stannard}}},\ }\href@noop {} {\emph {\bibinfo {title} {Relativity: A very
  short introduction}}}\ (\bibinfo  {publisher} {Oxford university press},\
  \bibinfo {address} {London},\ \bibinfo {year} {2008})\BibitemShut {NoStop}%
\bibitem [{\citenamefont {{Hartle}}(2002)}]{hartle_book}%
  \BibitemOpen
  \bibfield  {author} {\bibinfo {author} {\bibfnamefont {J.}~\bibnamefont
  {{Hartle}}},\ }\href@noop {} {\emph {\bibinfo {title} {Gravity: An
  introduction to Einstein's General Relativity}}}\ (\bibinfo  {publisher}
  {Pearson},\ \bibinfo {address} {California},\ \bibinfo {year}
  {2002})\BibitemShut {NoStop}%
\bibitem [{\citenamefont {{Misner}}(1973)}]{wheeler_book}%
  \BibitemOpen
  \bibfield  {author} {\bibinfo {author} {\bibfnamefont {T.~K. W. J.~A.}\
  \bibnamefont {{Misner}}, \bibfnamefont {James}},\ }\href@noop {} {\emph
  {\bibinfo {title} {Gravitation}}}\ (\bibinfo  {publisher} {Palgrave
  Macmillan},\ \bibinfo {address} {USA},\ \bibinfo {year} {1973})\BibitemShut
  {NoStop}%
\bibitem [{\citenamefont {Tolman}(1939)}]{Tolman39}%
  \BibitemOpen
  \bibfield  {author} {\bibinfo {author} {\bibfnamefont {R.~C.}\ \bibnamefont
  {Tolman}},\ }\href@noop {} {\bibfield  {journal} {\bibinfo  {journal} {Phys.
  Rev.}\ }\textbf {\bibinfo {volume} {55}},\ \bibinfo {pages} {364} (\bibinfo
  {year} {1939})}\BibitemShut {NoStop}%
\bibitem [{Note1()}]{Note1}%
  \BibitemOpen
  \bibinfo {note} {Note that no electrical or magnetical fields are
  considered.}\BibitemShut {Stop}%
\bibitem [{\citenamefont {Radice}\ \emph {et~al.}(2018)\citenamefont {Radice},
  \citenamefont {Perego}, \citenamefont {Zappa},\ and\ \citenamefont
  {Bernuzzi}}]{Radice:2017lry}%
  \BibitemOpen
  \bibfield  {author} {\bibinfo {author} {\bibfnamefont {D.}~\bibnamefont
  {Radice}}, \bibinfo {author} {\bibfnamefont {A.}~\bibnamefont {Perego}},
  \bibinfo {author} {\bibfnamefont {F.}~\bibnamefont {Zappa}}, \ and\ \bibinfo
  {author} {\bibfnamefont {S.}~\bibnamefont {Bernuzzi}},\ }\href {\doibase
  10.3847/2041-8213/aaa402} {\bibfield  {journal} {\bibinfo  {journal}
  {Astrophys. J.}\ }\textbf {\bibinfo {volume} {852}},\ \bibinfo {pages} {L29}
  (\bibinfo {year} {2018})},\ \Eprint {http://arxiv.org/abs/1711.03647}
  {arXiv:1711.03647 [astro-ph.HE]} \BibitemShut {NoStop}%
\bibitem [{\citenamefont {Margalit}\ and\ \citenamefont
  {Metzger}(2017)}]{Margalit:2017dij}%
  \BibitemOpen
  \bibfield  {author} {\bibinfo {author} {\bibfnamefont {B.}~\bibnamefont
  {Margalit}}\ and\ \bibinfo {author} {\bibfnamefont {B.~D.}\ \bibnamefont
  {Metzger}},\ }\href {\doibase 10.3847/2041-8213/aa991c} {\bibfield  {journal}
  {\bibinfo  {journal} {Astrophys. J.}\ }\textbf {\bibinfo {volume} {850}},\
  \bibinfo {pages} {L19} (\bibinfo {year} {2017})},\ \Eprint
  {http://arxiv.org/abs/1710.05938} {arXiv:1710.05938 [astro-ph.HE]}
  \BibitemShut {NoStop}%
\bibitem [{\citenamefont {Rezzolla}\ \emph {et~al.}(2018)\citenamefont
  {Rezzolla}, \citenamefont {Most},\ and\ \citenamefont
  {Weih}}]{Rezzolla:2017aly}%
  \BibitemOpen
  \bibfield  {author} {\bibinfo {author} {\bibfnamefont {L.}~\bibnamefont
  {Rezzolla}}, \bibinfo {author} {\bibfnamefont {E.~R.}\ \bibnamefont {Most}},
  \ and\ \bibinfo {author} {\bibfnamefont {L.~R.}\ \bibnamefont {Weih}},\
  }\href {\doibase 10.3847/2041-8213/aaa401} {\bibfield  {journal} {\bibinfo
  {journal} {Astrophys. J.}\ }\textbf {\bibinfo {volume} {852}},\ \bibinfo
  {pages} {L25} (\bibinfo {year} {2018})},\ \bibinfo {note} {[Astrophys. J.
  Lett.852,L25(2018)]},\ \Eprint {http://arxiv.org/abs/1711.00314}
  {arXiv:1711.00314 [astro-ph.HE]} \BibitemShut {NoStop}%
\bibitem [{\citenamefont {Abbott}\ \emph {et~al.}(2018)\citenamefont {Abbott}
  \emph {et~al.}}]{Abbott:2018exr}%
  \BibitemOpen
  \bibfield  {author} {\bibinfo {author} {\bibfnamefont {B.~P.}\ \bibnamefont
  {Abbott}} \emph {et~al.} (\bibinfo {collaboration} {Virgo, LIGO
  Scientific}),\ }\href@noop {} {\  (\bibinfo {year} {2018})},\ \Eprint
  {http://arxiv.org/abs/1805.11581} {arXiv:1805.11581 [gr-qc]} \BibitemShut
  {NoStop}%
\bibitem [{\citenamefont {Most}\ \emph {et~al.}(2018)\citenamefont {Most},
  \citenamefont {Weih}, \citenamefont {Rezzolla},\ and\ \citenamefont
  {Schaffner-Bielich}}]{Most:2018hfd}%
  \BibitemOpen
  \bibfield  {author} {\bibinfo {author} {\bibfnamefont {E.~R.}\ \bibnamefont
  {Most}}, \bibinfo {author} {\bibfnamefont {L.~R.}\ \bibnamefont {Weih}},
  \bibinfo {author} {\bibfnamefont {L.}~\bibnamefont {Rezzolla}}, \ and\
  \bibinfo {author} {\bibfnamefont {J.}~\bibnamefont {Schaffner-Bielich}},\
  }\href {\doibase 10.1103/PhysRevLett.120.261103} {\bibfield  {journal}
  {\bibinfo  {journal} {Phys. Rev. Lett.}\ }\textbf {\bibinfo {volume} {120}},\
  \bibinfo {pages} {261103} (\bibinfo {year} {2018})},\ \Eprint
  {http://arxiv.org/abs/1803.00549} {arXiv:1803.00549 [gr-qc]} \BibitemShut
  {NoStop}%
\bibitem [{\citenamefont {De}\ \emph {et~al.}(2018)\citenamefont {De},
  \citenamefont {Finstad}, \citenamefont {Lattimer}, \citenamefont {Brown},
  \citenamefont {Berger},\ and\ \citenamefont {Biwer}}]{De:2018uhw}%
  \BibitemOpen
  \bibfield  {author} {\bibinfo {author} {\bibfnamefont {S.}~\bibnamefont
  {De}}, \bibinfo {author} {\bibfnamefont {D.}~\bibnamefont {Finstad}},
  \bibinfo {author} {\bibfnamefont {J.~M.}\ \bibnamefont {Lattimer}}, \bibinfo
  {author} {\bibfnamefont {D.~A.}\ \bibnamefont {Brown}}, \bibinfo {author}
  {\bibfnamefont {E.}~\bibnamefont {Berger}}, \ and\ \bibinfo {author}
  {\bibfnamefont {C.~M.}\ \bibnamefont {Biwer}},\ }\href@noop {} {\  (\bibinfo
  {year} {2018})},\ \Eprint {http://arxiv.org/abs/1804.08583} {arXiv:1804.08583
  [astro-ph.HE]} \BibitemShut {NoStop}%
\bibitem [{\citenamefont {Kumar}\ \emph {et~al.}(2018)\citenamefont {Kumar},
  \citenamefont {Agrawal},\ and\ \citenamefont {Patra}}]{Kumar:2017wqp}%
  \BibitemOpen
  \bibfield  {author} {\bibinfo {author} {\bibfnamefont {B.}~\bibnamefont
  {Kumar}}, \bibinfo {author} {\bibfnamefont {B.~K.}\ \bibnamefont {Agrawal}},
  \ and\ \bibinfo {author} {\bibfnamefont {S.~K.}\ \bibnamefont {Patra}},\
  }\href {\doibase 10.1103/PhysRevC.97.045806} {\bibfield  {journal} {\bibinfo
  {journal} {Phys. Rev.}\ }\textbf {\bibinfo {volume} {C97}},\ \bibinfo {pages}
  {045806} (\bibinfo {year} {2018})},\ \Eprint
  {http://arxiv.org/abs/1711.04940} {arXiv:1711.04940 [nucl-th]} \BibitemShut
  {NoStop}%
\bibitem [{\citenamefont {Fattoyev}\ \emph {et~al.}(2018)\citenamefont
  {Fattoyev}, \citenamefont {Piekarewicz},\ and\ \citenamefont
  {Horowitz}}]{Fattoyev:2017jql}%
  \BibitemOpen
  \bibfield  {author} {\bibinfo {author} {\bibfnamefont {F.~J.}\ \bibnamefont
  {Fattoyev}}, \bibinfo {author} {\bibfnamefont {J.}~\bibnamefont
  {Piekarewicz}}, \ and\ \bibinfo {author} {\bibfnamefont {C.~J.}\ \bibnamefont
  {Horowitz}},\ }\href {\doibase 10.1103/PhysRevLett.120.172702} {\bibfield
  {journal} {\bibinfo  {journal} {Phys. Rev. Lett.}\ }\textbf {\bibinfo
  {volume} {120}},\ \bibinfo {pages} {172702} (\bibinfo {year} {2018})},\
  \Eprint {http://arxiv.org/abs/1711.06615} {arXiv:1711.06615 [nucl-th]}
  \BibitemShut {NoStop}%
\bibitem [{\citenamefont {Malik}\ \emph {et~al.}(2018)\citenamefont {Malik},
  \citenamefont {Alam}, \citenamefont {Fortin}, \citenamefont {Providencia},
  \citenamefont {Agrawal}, \citenamefont {Jha}, \citenamefont {Kumar},\ and\
  \citenamefont {Patra}}]{Malik:2018zcf}%
  \BibitemOpen
  \bibfield  {author} {\bibinfo {author} {\bibfnamefont {T.}~\bibnamefont
  {Malik}}, \bibinfo {author} {\bibfnamefont {N.}~\bibnamefont {Alam}},
  \bibinfo {author} {\bibfnamefont {M.}~\bibnamefont {Fortin}}, \bibinfo
  {author} {\bibfnamefont {C.}~\bibnamefont {Providencia}}, \bibinfo {author}
  {\bibfnamefont {B.~K.}\ \bibnamefont {Agrawal}}, \bibinfo {author}
  {\bibfnamefont {T.~K.}\ \bibnamefont {Jha}}, \bibinfo {author} {\bibfnamefont
  {B.}~\bibnamefont {Kumar}}, \ and\ \bibinfo {author} {\bibfnamefont {S.~K.}\
  \bibnamefont {Patra}},\ }\href@noop {} {\  (\bibinfo {year} {2018})},\
  \Eprint {http://arxiv.org/abs/1805.11963} {arXiv:1805.11963 [nucl-th]}
  \BibitemShut {NoStop}%
\bibitem [{\citenamefont {Hornick}\ \emph {et~al.}(2018)\citenamefont
  {Hornick}, \citenamefont {Tolos}, \citenamefont {Zacchi}, \citenamefont
  {Christian},\ and\ \citenamefont {Schaffner-Bielich}}]{Hornick:2018kfi}%
  \BibitemOpen
  \bibfield  {author} {\bibinfo {author} {\bibfnamefont {N.}~\bibnamefont
  {Hornick}}, \bibinfo {author} {\bibfnamefont {L.}~\bibnamefont {Tolos}},
  \bibinfo {author} {\bibfnamefont {A.}~\bibnamefont {Zacchi}}, \bibinfo
  {author} {\bibfnamefont {J.-E.}\ \bibnamefont {Christian}}, \ and\ \bibinfo
  {author} {\bibfnamefont {J.}~\bibnamefont {Schaffner-Bielich}},\ }\href
  {\doibase 10.1103/PhysRevC.98.065804} {\bibfield  {journal} {\bibinfo
  {journal} {Phys. Rev.}\ }\textbf {\bibinfo {volume} {C98}},\ \bibinfo {pages}
  {065804} (\bibinfo {year} {2018})},\ \Eprint
  {http://arxiv.org/abs/1808.06808} {arXiv:1808.06808 [astro-ph.HE]}
  \BibitemShut {NoStop}%
\bibitem [{\citenamefont {{Thorne}}\ and\ \citenamefont
  {{Campolattaro}}(1967)}]{1967ApJ...149..591T}%
  \BibitemOpen
  \bibfield  {author} {\bibinfo {author} {\bibfnamefont {K.~S.}\ \bibnamefont
  {{Thorne}}}\ and\ \bibinfo {author} {\bibfnamefont {A.}~\bibnamefont
  {{Campolattaro}}},\ }\href {\doibase 10.1086/149288} {\bibfield  {journal}
  {\bibinfo  {journal} {\apj}\ }\textbf {\bibinfo {volume} {149}},\ \bibinfo
  {pages} {591} (\bibinfo {year} {1967})}\BibitemShut {NoStop}%
\bibitem [{\citenamefont {Baym}\ and\ \citenamefont
  {Chin}(1976)}]{Baym:1976yu}%
  \BibitemOpen
  \bibfield  {author} {\bibinfo {author} {\bibfnamefont {G.}~\bibnamefont
  {Baym}}\ and\ \bibinfo {author} {\bibfnamefont {S.~A.}\ \bibnamefont
  {Chin}},\ }\href@noop {} {\bibfield  {journal} {\bibinfo  {journal} {Phys.
  Lett.}\ }\textbf {\bibinfo {volume} {B62}},\ \bibinfo {pages} {241} (\bibinfo
  {year} {1976})}\BibitemShut {NoStop}%
\bibitem [{\citenamefont {Aerts}\ \emph {et~al.}(1978)\citenamefont {Aerts},
  \citenamefont {Mulders},\ and\ \citenamefont {de~Swart}}]{Aerts78}%
  \BibitemOpen
  \bibfield  {author} {\bibinfo {author} {\bibfnamefont {A.~T.~M.}\
  \bibnamefont {Aerts}}, \bibinfo {author} {\bibfnamefont {P.~J.~G.}\
  \bibnamefont {Mulders}}, \ and\ \bibinfo {author} {\bibfnamefont {J.~J.}\
  \bibnamefont {de~Swart}},\ }\href@noop {} {\bibfield  {journal} {\bibinfo
  {journal} {Phys. Rev. D}\ }\textbf {\bibinfo {volume} {17}},\ \bibinfo
  {pages} {260} (\bibinfo {year} {1978})}\BibitemShut {NoStop}%
\bibitem [{\citenamefont {Schertler}\ \emph {et~al.}(1998)\citenamefont
  {Schertler}, \citenamefont {Greiner},\ and\ \citenamefont
  {Thoma}}]{Schertler:1998cs}%
  \BibitemOpen
  \bibfield  {author} {\bibinfo {author} {\bibfnamefont {K.}~\bibnamefont
  {Schertler}}, \bibinfo {author} {\bibfnamefont {C.}~\bibnamefont {Greiner}},
  \ and\ \bibinfo {author} {\bibfnamefont {M.}~\bibnamefont {Thoma}},\
  }\href@noop {} {\  (\bibinfo {year} {1998})},\ \Eprint
  {http://arxiv.org/abs/astro-ph/9801200} {arXiv:astro-ph/9801200 [astro-ph]}
  \BibitemShut {NoStop}%
\bibitem [{\citenamefont {Klahn}\ and\ \citenamefont
  {Fischer}(2015)}]{Klahn:2015mfa}%
  \BibitemOpen
  \bibfield  {author} {\bibinfo {author} {\bibfnamefont {T.}~\bibnamefont
  {Klahn}}\ and\ \bibinfo {author} {\bibfnamefont {T.}~\bibnamefont
  {Fischer}},\ }\href@noop {} {\  (\bibinfo {year} {2015})},\ \Eprint
  {http://arxiv.org/abs/1503.07442} {arXiv:1503.07442 [nucl-th]} \BibitemShut
  {NoStop}%
\bibitem [{\citenamefont {Raaijmakers}\ \emph {et~al.}(2019)\citenamefont
  {Raaijmakers} \emph {et~al.}}]{Raaijmakers:2019qny}%
  \BibitemOpen
  \bibfield  {author} {\bibinfo {author} {\bibfnamefont {G.}~\bibnamefont
  {Raaijmakers}} \emph {et~al.},\ }\href {\doibase 10.3847/2041-8213/ab451a}
  {\bibfield  {journal} {\bibinfo  {journal} {Astrophys. J.}\ }\textbf
  {\bibinfo {volume} {887}},\ \bibinfo {pages} {L22} (\bibinfo {year}
  {2019})},\ \Eprint {http://arxiv.org/abs/1912.05703} {arXiv:1912.05703
  [astro-ph.HE]} \BibitemShut {NoStop}%
\bibitem [{\citenamefont {Abbott}\ \emph
  {et~al.}(2019{\natexlab{b}})\citenamefont {Abbott} \emph
  {et~al.}}]{Abbott:2018qee}%
  \BibitemOpen
  \bibfield  {author} {\bibinfo {author} {\bibfnamefont {B.~P.}\ \bibnamefont
  {Abbott}} \emph {et~al.} (\bibinfo {collaboration} {LIGO Scientific,
  Virgo}),\ }\href {\doibase 10.3847/1538-4357/ab113b} {\bibfield  {journal}
  {\bibinfo  {journal} {Astrophys. J.}\ }\textbf {\bibinfo {volume} {875}},\
  \bibinfo {pages} {122} (\bibinfo {year} {2019}{\natexlab{b}})},\ \Eprint
  {http://arxiv.org/abs/1812.11656} {arXiv:1812.11656 [astro-ph.HE]}
  \BibitemShut {NoStop}%
\bibitem [{\citenamefont {Abbott}\ \emph {et~al.}(2020)\citenamefont {Abbott}
  \emph {et~al.}}]{Abbott:2020khf}%
  \BibitemOpen
  \bibfield  {author} {\bibinfo {author} {\bibfnamefont {R.}~\bibnamefont
  {Abbott}} \emph {et~al.} (\bibinfo {collaboration} {LIGO Scientific,
  Virgo}),\ }\href {\doibase 10.3847/2041-8213/ab960f} {\bibfield  {journal}
  {\bibinfo  {journal} {Astrophys. J.}\ }\textbf {\bibinfo {volume} {896}},\
  \bibinfo {pages} {L44} (\bibinfo {year} {2020})},\ \Eprint
  {http://arxiv.org/abs/2006.12611} {arXiv:2006.12611 [astro-ph.HE]}
  \BibitemShut {NoStop}%
\end{thebibliography}%
\bibliographystyle{apsrev4-1}
\end{document}